\documentclass[final,3p,times]{elsarticle}
\usepackage{amssymb}
\usepackage{amsmath}
\usepackage{lineno}

\newcommand*\patchAmsMathEnvironmentForLineno[1]{
  \expandafter\let\csname old#1\expandafter\endcsname\csname #1\endcsname
  \expandafter\let\csname oldend#1\expandafter\endcsname\csname end#1\endcsname
  \renewenvironment{#1}
     {\linenomath\csname old#1\endcsname}
     {\csname oldend#1\endcsname\endlinenomath}}
\newcommand*\patchBothAmsMathEnvironmentsForLineno[1]{
  \patchAmsMathEnvironmentForLineno{#1}
  \patchAmsMathEnvironmentForLineno{#1*}}
\AtBeginDocument{
\patchBothAmsMathEnvironmentsForLineno{equation}
\patchBothAmsMathEnvironmentsForLineno{align}
\patchBothAmsMathEnvironmentsForLineno{flalign}
\patchBothAmsMathEnvironmentsForLineno{alignat}
\patchBothAmsMathEnvironmentsForLineno{gather}
\patchBothAmsMathEnvironmentsForLineno{multline}
}

\usepackage{xcolor} % use \textcolor in revision
\usepackage{ulem} % use \sout in revision
\usepackage{cancel} % use \cancel for equations in revision
\usepackage{bm}
\usepackage{hyperref}
\biboptions{numbers,sort&compress} % [1,2,3,4] -> [1-4].

\journal{Journal of Computational Physics}

\begin{document}

\begin{frontmatter}



\title{Weakly Compressible Subcycling for Accelerating Simulations of Surface-Tension-Dominated Incompressible Two-Phase Flows}


\author[isct]{Shu Yamashita} %% Author name
\author[isct]{Shintaro Matsushita} %% Author name
\author[isct]{Tetsuya Suekane} %% Author name

\affiliation[isct]{organization={Institute of Science Tokyo, School of Engineering},%Department and Organization
            addressline={2-12-1, Ookayama, Meguro-ku}, 
            city={Tokyo},
            postcode={152-8550}, 
            country={Japan}}

%% Abstract
\begin{abstract}

Simulations of surface-tension-dominated incompressible two-phase flows are computationally expensive due to the severe capillary time-step constraint.
Although many studies have proposed time-implicit discretizations of surface tension to allow larger time-step sizes and accelerate simulations, these methods suffer from either artificial dissipation or complex implementation.
Here, we propose a simple and novel approach: an incompressible solver with weakly compressible subcycling.
The proposed approach relaxes the capillary time-step constraint, thereby accelerating simulations by more than $8.6\times$ without relying on artificially dissipative stabilization or requiring complex implementation.
The key idea is to introduce lightweight substeps using a weakly compressible solver to assist the main incompressible solver.
These substeps enable the main incompressible solver to use accurately computed fluxes and surface tension force, even with large time-step sizes.
Numerical tests demonstrate the effectiveness of the proposed approach for practical problems, including the Rayleigh--Plateau instability and two-phase flows in porous media.
This study paves the way for a new paradigm in which a weakly compressible solver serves as an assistant to an incompressible solver.

\end{abstract}


% \begin{highlights}
% \item Proposed an incompressible solver with weakly compressible subcycling.
% \item A weakly compressible solver serves as an assistant to the incompressible solver.
% \item The proposed incompressible solver relaxes the capillary time-step constraint.
% \item Accelerate simulations by more than 8.6 times in practical problems.
% \item The proposed solver avoids artificial dissipation and complex implementation.
% \end{highlights}

%% Keywords

\end{frontmatter}

% \linenumbers

\section{Introduction}

This study focuses on simulations of surface-tension-dominated incompressible two-phase flows.
These flows play a critical role in a wide range of applications, including fuel cells~\cite{LU20119864,ANDERSON20104531,YANG202315677}, water electrolysis for hydrogen production~\cite{Lettenmeier2017,Hodges2022,Kulkarni2023}, geological storage of CO$_2$~\cite{KREVOR2015221,ALI2022103895}, inkjet printing~\cite{He2017}, and microfluidics~\cite{Olanrewaju2018,JIANG2023116932}.
Numerical simulations are powerful and widely used tools for understanding and optimizing these applications. 
Therefore, improving the practicality of simulations further advances these fields.

For more practical simulations of surface-tension-dominated incompressible flows, reducing computational cost remains a major challenge.
Simulating these flows is computationally expensive because the time-explicit discretization of surface tension imposes a severe capillary time-step constraint~\cite{BRACKBILL1992335}.
This constraint requires small time-step sizes, thereby increasing the number of times the pressure Poisson equation is solved, which is a major computational cost in incompressible flow simulations~\cite{DODD2014416,alsalti2022poisson,FRANTZIS201928}.

Although many studies have developed time-implicit discretizations of surface tension to allow larger time-step sizes, these methods generally suffer from either artificial dissipation or implementation complexity. 
For example, Hysing~\cite{Hysing2006} proposed an implicit surface-tension treatment based on the finite element method. 
Raessi et al.~\cite{Raessi2009} extended this approach to the finite volume method and showed that the capillary time-step constraint could be exceeded by a factor of at least five while maintaining stability. 
Denner et al.~\cite{DENNER201759} and Popinet~\cite{Popinet2018} pointed out that the stabilization achieved by the implicit surface-tension treatments of Hysing~\cite{Hysing2006} and Raessi et al.~\cite{Raessi2009} is essentially due to artificial dissipation introduced near the interface.
Fahsi and Soulaïmani~\cite{FAHSI2026106889} reduced this artificial dissipation using a second-order time-integration scheme. 
However, the remaining artificial dissipation still degrades accuracy.
Denner et al.~\cite{DENNER2022111128} and Janodet et al.~\cite{JANODET2025113520} developed fully coupled algorithms.
These methods allow time-step sizes larger than the capillary time-step constraint without relying on artificially dissipative stabilization but require complex implementations, including linearizing the nonlinear surface tension term and solving a large coupled linear system involving the continuity, momentum, and volume-of-fluid advection equations.

As an alternative approach, Sussman and Ohta~\cite{Sussman2009,Sussman2012} used volume-preserving motion by mean curvature to model surface tension.
Their method relaxes the capillary time-step constraint but requires the level-set method, its associated reinitialization procedure, and the computation of the average curvature for each connected interface.

In this study, we propose a novel approach, an incompressible solver with \textbf{weakly compressible subcycling}, which is fundamentally different from time-implicit discretizations of surface tension.
Unlike existing methods, the proposed method relaxes the capillary time-step constraint without relying on artificially dissipative stabilization or requiring complex implementations.
The key idea is to introduce lightweight substeps using a weakly compressible solver to assist the main incompressible solver.
These substeps enable the main incompressible solver to use accurately computed fluxes and surface tension force, even with large time-step sizes.
To the best of our knowledge, this is the first study proposing the use of a weakly compressible solver to assist an incompressible solver.

\section{Preliminaries: Standard Incompressible Solver for Two-Phase Flows}
\label{sec:standard}

This section presents a standard incompressible solver for two-phase flows, which serves as the baseline for the proposed incompressible solver with weakly compressible subcycling, presented in the next section.
In the last part of this section (Section~\ref{sec:capillary_time_step_constraint}), we provide an intuitive explanation of the capillary time-step constraint that limits the time-step size of the standard solver.
This explanation helps readers understand the key idea underlying the proposed method.

\subsection{Governing Equations}
The Navier--Stokes equations govern the dynamics of incompressible two-phase flows with surface tension:
\begin{align}
    \frac{\partial (\rho \bm u)}{\partial t} + \nabla \cdot (\rho \bm u \bm u) &= -\nabla p + \nabla \cdot \mu \left[ \nabla \bm u + \left(\nabla \bm u\right)^\top \right] + \bm f_\sigma, \label{eq:NS} \\
    \nabla \cdot \bm u &= 0, \label{eq:continuity}
\end{align}
where $\rho$ is the density, $\bm u = (u, v, w)$ is the velocity, $p$ is the pressure, $\mu$ is the viscosity, and $\bm f_\sigma = ({f_\sigma}_x, {f_\sigma}_y, {f_\sigma}_z)$ is the surface tension force.
For simplicity, gravity is neglected.

For the interface-capturing method, we use the phase-field method, specifically the accurate conservative diffuse-interface (ACDI) method~\cite{JAIN2022111529}.
In the ACDI method, the phase at position $\bm x$ is represented by the phase-field variable $\phi(\bm x)$, which takes values in $[0,1]$.
$\phi(\bm x) \approx 0$ indicates phase~1, whereas $\phi(\bm x) \approx 1$ indicates phase~2.
The interface is represented by a smooth transition region spanning several grid cells.
The evolution of the phase-field variable is governed by
\begin{equation}
    \frac{\partial \phi}{\partial t} + \nabla \cdot (\bm u \phi) = \nabla \cdot \left\{ \Gamma \left\{ \epsilon \nabla \phi - \frac{1}{4} \left[ 1 - \tanh^2\left(\frac{\psi}{2 \epsilon}\right) \right] \frac{\nabla \psi}{\| \nabla \psi \|} \right\} \right\}, \label{eq:ACDI}
\end{equation}
where $\Gamma$ is a positive parameter that controls how strongly the right-hand side of Eq.~(\ref{eq:ACDI}) preserves the phase-field profile at the interface, and $\epsilon$ is a positive parameter that controls the interface thickness.
We set $\Gamma=\|\bm u\|_{\mathrm{max}}$ and $\epsilon=\Delta x$ as in~\cite{MIRJALILI2019221}, where $\|\bm u\|_{\mathrm{max}}$ denotes the maximum velocity magnitude over the domain.
The auxiliary signed-distance-like variable $\psi$ is given by
$\psi = \epsilon \ln \left[\phi / (1 - \phi)\right]$.

The density and viscosity are computed as $\rho = \rho_1 (1 - \phi) + \rho_2 \phi$ and $\mu = \mu_1 (1 - \phi) + \mu_2 \phi$, where subscripts $1$ and $2$ denote phases~1 and~2, respectively.
To model the surface tension force, we use the localized continuum surface force model~\cite{MIRJALILI2023111795}, $\bm f_\sigma =  6 \phi (1 - \phi) \sigma \kappa \nabla \phi$, where $\sigma$ is the surface tension coefficient and $\kappa = - \nabla \cdot \left(\nabla \psi / \|\nabla \psi\|\right)$ is the interface curvature.

\subsection{Numerical Methods}
Here, we present an overview of the numerical procedure for advancing the solution from time step $n$ to $n+1$; the detailed implementation is described in \ref{sec:detailed_impl_standard}.
For simplicity, the following description is given for the 2D case; the extension to 3D is straightforward.
For convenience, we rewrite Eqs.~(\ref{eq:NS}) and~(\ref{eq:ACDI}) in terms of the momentum fluxes $\bm F_{\rho u}$ and $\bm F_{\rho v}$, which include advective and viscous contributions, and the volume flux $\bm F_\phi$:
\begin{align}
    \frac{\partial (\rho u)}{\partial t} &= \nabla \cdot \bm F_{\rho u} - \frac{\partial p}{\partial x} + {f_\sigma}_x, 
    \quad &\text{where}& \quad
    \bm F_{\rho u} = - \rho u \bm u + \mu \left( 2 \frac{\partial u}{\partial x},\ \frac{\partial u}{\partial y} + \frac{\partial v}{\partial x}\right), \\
    \frac{\partial (\rho v)}{\partial t} &= \nabla \cdot \bm F_{\rho v} - \frac{\partial p}{\partial y} + {f_\sigma}_y, 
    \quad &\text{where}& \quad
    \bm F_{\rho v} = - \rho v \bm u + \mu \left(\frac{\partial u}{\partial y} + \frac{\partial v}{\partial x},\ 2\frac{\partial v}{\partial y}\right), \\
    \frac{\partial \phi}{\partial t} &= \nabla \cdot \bm F_\phi,
    \quad &\text{where}& \quad
    \bm F_{\phi} = - \bm u \phi + \Gamma \left\{ \epsilon \nabla \phi - \frac{1}{4} \left[ 1 - \tanh^2\left(\frac{\psi}{2 \epsilon}\right) \right] \frac{\nabla \psi}{\| \nabla \psi \|} \right\}.
\end{align}
We employ the projection method~\cite{chorin1968projection}.
The governing equations are discretized with the finite-volume method on a uniform Cartesian grid, and time integration is performed with the first-order explicit Euler method.
All scalar variables ($\phi$, $\rho$, $\mu$, and $p$) are stored at cell centers, whereas the velocity is stored at both cell centers ($\bm u_c$) and cell faces ($\bm u_f$).
In what follows, $c \rightarrow f$ and $f \rightarrow c$ denote linear interpolation from cell centers to cell faces and from cell faces to cell centers, respectively.
The numerical procedure for advancing the solution from time step $n$ to $n+1$ is as follows:
\begin{enumerate}
    \item Compute the volume flux $\bm F_\phi$ and the momentum fluxes $\bm F_{\rho u}$ and $\bm F_{\rho v}$ at cell faces using the variables at time step $n$.
    The momentum fluxes are computed to be consistent with the volume flux~\cite{Yang2022}.
    \item Compute $\phi$ at time step $n + 1$: $\phi^{n + 1} = \phi^n + \nabla \cdot \bm F_\phi \Delta t$.
    \item Compute the intermediate cell-center velocity: $\bm u_c^* = \left(\rho^n u_c^n + \nabla \cdot \bm F_{\rho u}\Delta t,\ \rho^n v_c^n + \nabla \cdot \bm F_{\rho v}\Delta t\right) / \rho^{n+1}$.
    \item Compute the intermediate cell-face velocity $\bm u_f^*$ by linearly interpolating $\bm u_c^*$.
    \item Update the intermediate velocities $\bm u_f^*$ and $\bm u_c^*$ using the surface tension force: $\bm u_f^* \leftarrow \bm u_f^* + \bm f_\sigma \Delta t / \rho_{c \rightarrow f}^{n+1}$ and $\bm u_c^* \leftarrow \bm u_c^{*} + \left(\bm f_\sigma \Delta t / \rho_{c \rightarrow f}^{n+1}\right)_{f \rightarrow c}$, where the surface tension force $\bm f_\sigma = 6 \phi^{n + 1} (1 - \phi^{n + 1}) \sigma \kappa^{n + 1} \nabla \phi^{n + 1}$ is computed at cell faces.
    \item Solve the pressure Poisson equation: $\nabla \cdot \left(\nabla p^{n + 1} / \rho_{c \rightarrow f}^{n + 1}\right) = \nabla \cdot \bm u_f^* / \Delta t$.
    We solve this equation as a linear system by first applying diagonal scaling and then using the FlexGMRES solver with the PFMG preconditioner provided by the \textit{hypre} library~\cite{hypre, Falgout2002}, with a convergence tolerance of $10^{-7}$.
    \item Compute the velocities at time step $n+1$: $\bm u_f^{n + 1} = \bm u_f^* - \nabla p^{n + 1} \Delta t / \rho_{c \rightarrow f}^{n + 1}$ and $\bm u_c^{n + 1} = \bm u_c^* - \left(\nabla p^{n + 1} / \rho_{c \rightarrow f}^{n + 1}\right)_{f \rightarrow c} \Delta t$.
\end{enumerate}

\subsection{Intuitive Explanation of the Capillary Time-Step Constraint}
\label{sec:capillary_time_step_constraint}

As we demonstrate later in Section~\ref{sec:drop_oscillation} (droplet oscillation test), the above standard incompressible solver cannot significantly exceed the capillary time-step constraint~\cite{BRACKBILL1992335}:
\begin{equation}
    \Delta t \leq \sqrt{\frac{(\rho_1 + \rho_2) \Delta h^3}{4 \pi \sigma}} \equiv \Delta t_\sigma, 
    \label{eq:capillary_time_step_constraint}
\end{equation}
where $\Delta h = \min(\Delta x, \Delta y, \Delta z)$ is the minimum grid spacing.
Here, we provide an intuitive explanation of why using time-step sizes larger than $\Delta t_\sigma$ leads to inaccurate or unstable simulations.
This explanation helps readers understand the key idea of the proposed method presented in Section~\ref{sec:subcycling}.

The primary reason for inaccurate or unstable simulations is that large time-step sizes prevent accurate computation of the fluxes and the surface tension force.
The capillary time-step constraint in Eq.~\ref{eq:capillary_time_step_constraint} ensures that the time-step size is sufficiently small to temporally resolve the propagation of capillary waves~\cite{DENNER2015timestep}.
More specifically, even the fastest wave resolvable on the grid, with a wavelength of $2\Delta h$, propagates by no more than $\Delta h/2$ (i.e., one-quarter of its wavelength) during a single time step~\cite{BRACKBILL1992335}.
As a result, each cell can accurately compute the fluxes and the surface tension force because these quantities do not change significantly over such a small time step.
However, excessively large time-step sizes allow the fluxes and the surface tension force to change significantly within a single time step; thus, the time-explicit discretization cannot accurately compute these quantities, leading to inaccurate or unstable simulations.

\section{Proposed Method: Incompressible Solver with Weakly Compressible Subcycling}
\label{sec:subcycling}

\begin{figure}[!t]
    \centering
    \includegraphics[width=\linewidth]{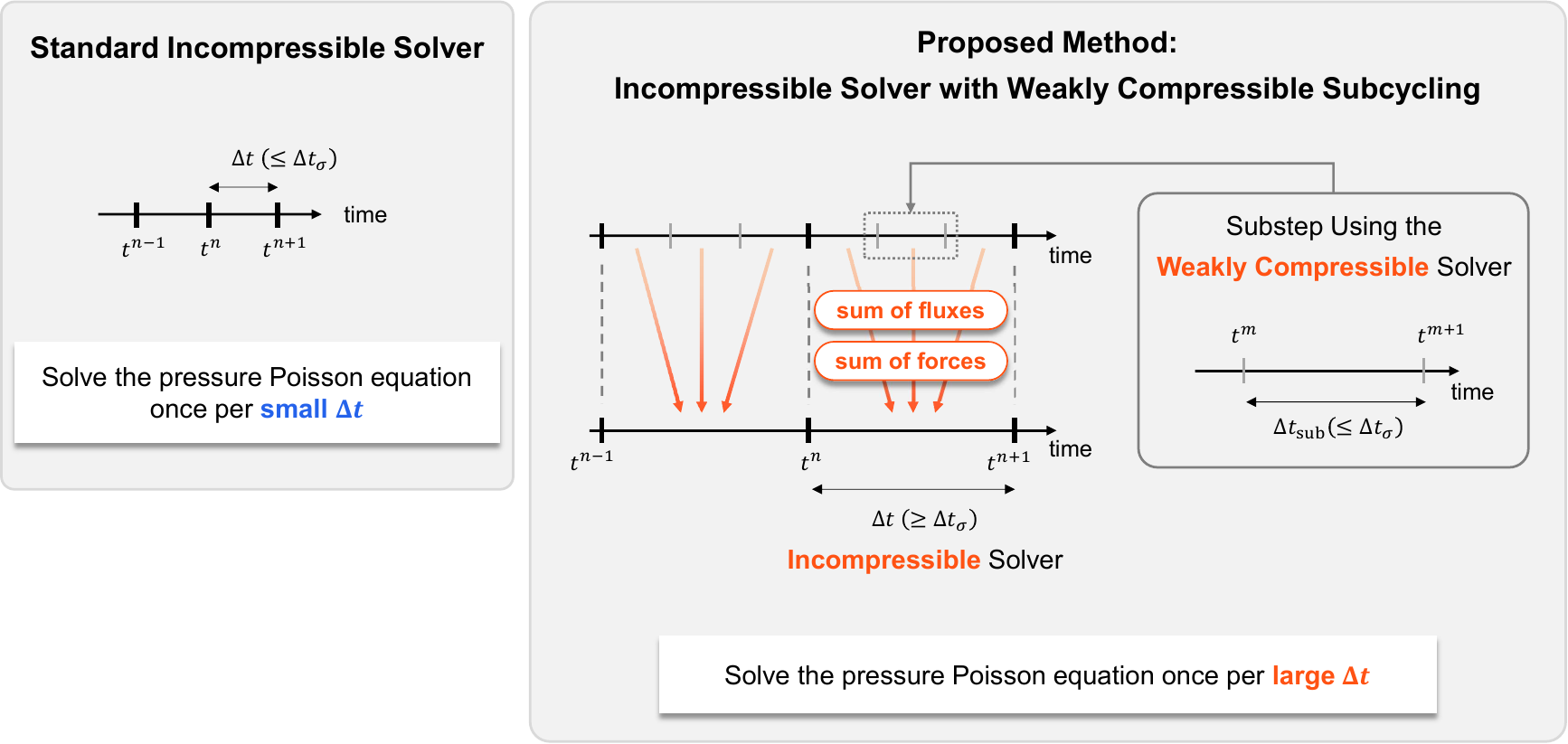}
    \caption{Overview of the proposed incompressible solver with weakly compressible subcycling. The lightweight weakly compressible solver assists the main incompressible solver by computing the fluxes and the surface tension force within each substep. This enables the main incompressible solver to exceed the capillary time-step constraint.}
    \label{fig:overview_subcycling}
\end{figure}

Our goal is to develop an incompressible solver that relaxes the capillary time-step constraint without relying on artificially dissipative stabilization or requiring complex implementation.
Using larger time-step sizes reduces the number of times the computationally expensive pressure Poisson equation is solved, thereby accelerating simulations.
In this section, we propose a novel approach: an incompressible solver with \textbf{weakly compressible subcycling}.
Figure~\ref{fig:overview_subcycling} provides an overview of the algorithm.
The key idea is to use the lightweight weakly compressible solver to assist the incompressible solver: the fluxes $\bm F_\phi$, $\bm F_{\rho u}$, and $\bm F_{\rho v}$ and the surface tension force $\bm f_\sigma$ required by the incompressible solver are computed by the weakly compressible solver in substeps.
As discussed in Section~\ref{sec:capillary_time_step_constraint}, enabling an incompressible solver to use large time-step sizes requires resolving capillary waves in time and, consequently, accurately evaluating the fluxes and the surface tension force acting during a single time step.
The proposed method achieves this by temporally resolving the propagation of capillary waves through small substeps within each time step.

We present the algorithm from time step $n$ to $n+1$.
The incompressible solver with weakly compressible subcycling comprises the following two stages:
\begin{enumerate}
    \item \textbf{Weakly compressible subcycling.}
    From time $t^n$ to $t^{n+1}$, we compute the evolution of the flow (i.e., $\bm{u}$, $p$, and $\phi$) using multiple substeps, each of which is solved with a weakly compressible solver. 
    This weakly compressible solver is a lightweight solver that avoids solving the pressure Poisson equation by allowing slight compressibility. 
    In this study, we employ a weakly compressible solver based on the evolving pressure projection (EPP) method~\cite{YANG2021110113, Yang2022}. 
    The EPP method is similar to the standard incompressible solver in Section~\ref{sec:standard}, except for the pressure computation.
    Please refer to \ref{sec:detailed_impl_standard} for the detailed implementation of the common parts.
    The overview of the algorithm from substep $m$ to $m + 1$ is as follows:
    \begin{enumerate}
        \item Set the substep size $\Delta t_\mathrm{sub} = \min \left(t^{n + 1} - t^m, 0.95 \Delta t_u,\ 0.5 \Delta t_\mu,\ \Delta t_\sigma,\ 0.5 \Delta t_\phi\right)$, where $\Delta t_u = \Delta h / \| \bm u_f^m \|_\mathrm{max}$, $\Delta t_\mu = \Delta h^2/[2 d \max(\mu_1 / \rho_1,\ \mu_2 / \rho_2)]$, and $\Delta t_\phi = \Delta h^2/(2 d \Gamma \epsilon)$ are the time-step constraints for advection, viscosity, and the phase-field method, respectively. $d$ denotes the spatial dimension.
        \item Compute the fluxes $\bm F_\phi$, $\bm F_{\rho u}$, and $\bm F_{\rho v}$ using the variables at substep $m$.
        \item Compute $\phi$ at substep $m + 1$: $\phi^{m + 1} = \phi^m + \nabla \cdot \bm F_\phi \Delta t_\mathrm{sub}$.
        \item Compute the intermediate cell-center velocity: $\bm u_c^* = \left(\rho^m u_c^m + \nabla \cdot \bm F_{\rho u}\Delta t_\mathrm{sub},\ \rho^m v_c^m + \nabla \cdot \bm F_{\rho v}\Delta t_\mathrm{sub}\right) / \rho^{m+1}$.
        \item Compute the intermediate cell-face velocity $\bm u_f^*$ by linearly interpolating $\bm u_c^*$.
        \item Update the intermediate velocities $\bm u_f^*$ and $\bm u_c^*$ using the surface tension force: $\bm u_f^* \leftarrow \bm u_f^* + \bm f_\sigma \Delta t_\mathrm{sub} / \rho_{c \rightarrow f}^{m+1}$ and $\bm u_c^* \leftarrow \bm u_c^{*} + \left(\bm f_\sigma \Delta t_\mathrm{sub} / \rho_{c \rightarrow f}^{m+1}\right)_{f \rightarrow c}$, where the surface tension force $\bm f_\sigma = 6 \phi^{m + 1} (1 - \phi^{m + 1}) \sigma \kappa^{m + 1} \nabla \phi^{m + 1}$ is computed at cell faces.
        \item Iteratively update the pressure by solving the pressure evolution equation~\cite{Yang2022}:
        \begin{equation}
            \frac{p^{*,\ i + 1} - p^{*,\ i}}{\Delta t_\mathrm{sub}} + \rho^{m + 1} c_s^2 \nabla \cdot \left( \bm u_f^* - \frac{1}{\rho_{c \rightarrow f}^{m + 1}} \nabla p^{*,\ i} \Delta t_\mathrm{sub} \right) = 0,
        \end{equation}
        where $c_s$ is an artificial sound speed, which we set to $c_s = 0.4 \Delta h / \Delta t_\mathrm{sub}$. 
        The superscript $i~(0 \leq i < N_\mathrm{EPP})$ denotes the iteration count, and $p^{*,\ i}$ denotes the pressure at the $i$-th iteration.
        The iteration starts from $p^{*,\ 0} = p^m$ and ends with $p^{m+1} = p^{*,\ N_\mathrm{EPP}}$. 
        Increasing the number of iterations $N_\mathrm{EPP}$ mitigates compressibility~\cite{YANG2021110113}.
        We set $N_\mathrm{EPP} = 10$ based on our numerical experiments.
        This iteration is much less expensive than solving the pressure Poisson equation because of its simplicity, even though it does not enforce incompressibility.
        \item Compute the velocities at substep $m+1$: $\bm u_f^{m + 1} = \bm u_f^* - \nabla p^{m + 1} \Delta t_\mathrm{sub} / \rho_{c \rightarrow f}^{m + 1}$ and $\bm u_c^{m + 1} = \bm u_c^* - \left(\nabla p^{m + 1} / \rho_{c \rightarrow f}^{m + 1}\right)_{f \rightarrow c} \Delta t_\mathrm{sub}$.
    \end{enumerate}
    \item \textbf{Incompressible solver with fluxes and surface tension forces from subcycling.}
    After the above subcycling, we apply the standard incompressible solver from $t^n$ to $t^{n+1}$, replacing the fluxes and the surface tension force with their sums over all substeps:
    \begin{equation}
        \bm{F}_\phi \Delta t = \sum_m \left(\bm{F}_\phi \Delta t_\mathrm{sub}\right)^m, \quad
        \bm{F}_{\rho u} \Delta t = \sum_m \left(\bm{F}_{\rho u} \Delta t_\mathrm{sub}\right)^m, \quad
        \bm{F}_{\rho v} \Delta t = \sum_m \left(\bm{F}_{\rho v} \Delta t_\mathrm{sub}\right)^m, \quad
        \bm{f}_\sigma \Delta t = \sum_m \left(\bm{f}_\sigma \Delta t_\mathrm{sub}\right)^m.
    \end{equation}
\end{enumerate}
Notably, the implementation is simple. 
Both stages can be implemented by reusing the standard incompressible solver, with slight modifications to the pressure computation in stage 1 and to the computation of the fluxes and the surface tension force in stage 2.

\section{Numerical Tests}

\subsection{Inviscid 2D Droplet Oscillation}
\label{sec:drop_oscillation}

\paragraph{Objective}
We present simulations of an inviscid droplet oscillating due to surface tension.
The objective is to demonstrate that the proposed incompressible solver with weakly compressible subcycling can relax the capillary time-step constraint without relying on artificially dissipative stabilization, unlike existing dissipative approaches~\cite{Hysing2006,Raessi2009,DENNER201759,FAHSI2026106889}.

\paragraph{Problem Setup}
\begin{figure}[!t]
    \centering
    \includegraphics[width=\linewidth]{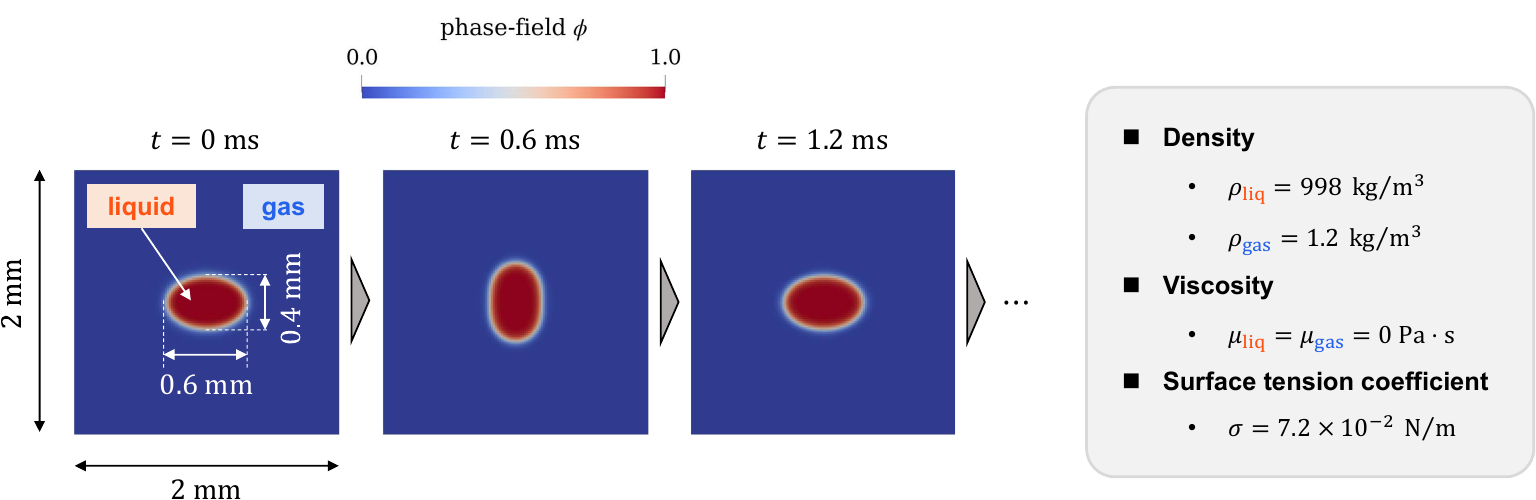}
    \caption{Inviscid 2D droplet oscillation test in Section~\ref{sec:drop_oscillation}.}
    \label{fig:drop_oscillation:schematic}
\end{figure}
Figure~\ref{fig:drop_oscillation:schematic} illustrates the setup for this test.
An elliptical liquid droplet is initially positioned at the center of a square domain, $[-1,1]~\mathrm{mm} \times [-1,1]~\mathrm{mm}$, and is surrounded by gas.
The liquid and gas densities are $\rho_\mathrm{liq} = 998~\mathrm{kg/m^3}$ and $\rho_\mathrm{gas} = 1.2~\mathrm{kg/m^3}$, respectively.
Both phases are inviscid, and the surface tension coefficient is $\sigma = 7.2 \times 10^{-2}~\mathrm{N/m}$.
Gravity is ignored.
Initially, the velocity and pressure are set to $\bm u = \bm 0$ and $p = 0$, respectively, and the interface is given by $x^2/a^2 + y^2/b^2 = 1$ with $a = 0.3~\mathrm{mm}$ and $b = 0.2~\mathrm{mm}$.
Free-slip boundary conditions are applied to all boundaries.
The grid resolution is $128 \times 128$.
Under these settings, the dominant time-step constraints are the capillary ($\Delta t_\sigma \approx 2.1 \times 10^{-6}~\mathrm{s}$) and the phase-field ($\min_t \Delta t_\phi \approx 1.2 \times 10^{-5}~\mathrm{s}$), where $\min_t$ denotes the minimum over the entire simulation time.

\paragraph{Results (Kinetic Energy)}
\begin{figure}[!t]
    \centering
    \includegraphics[width=0.7\linewidth]{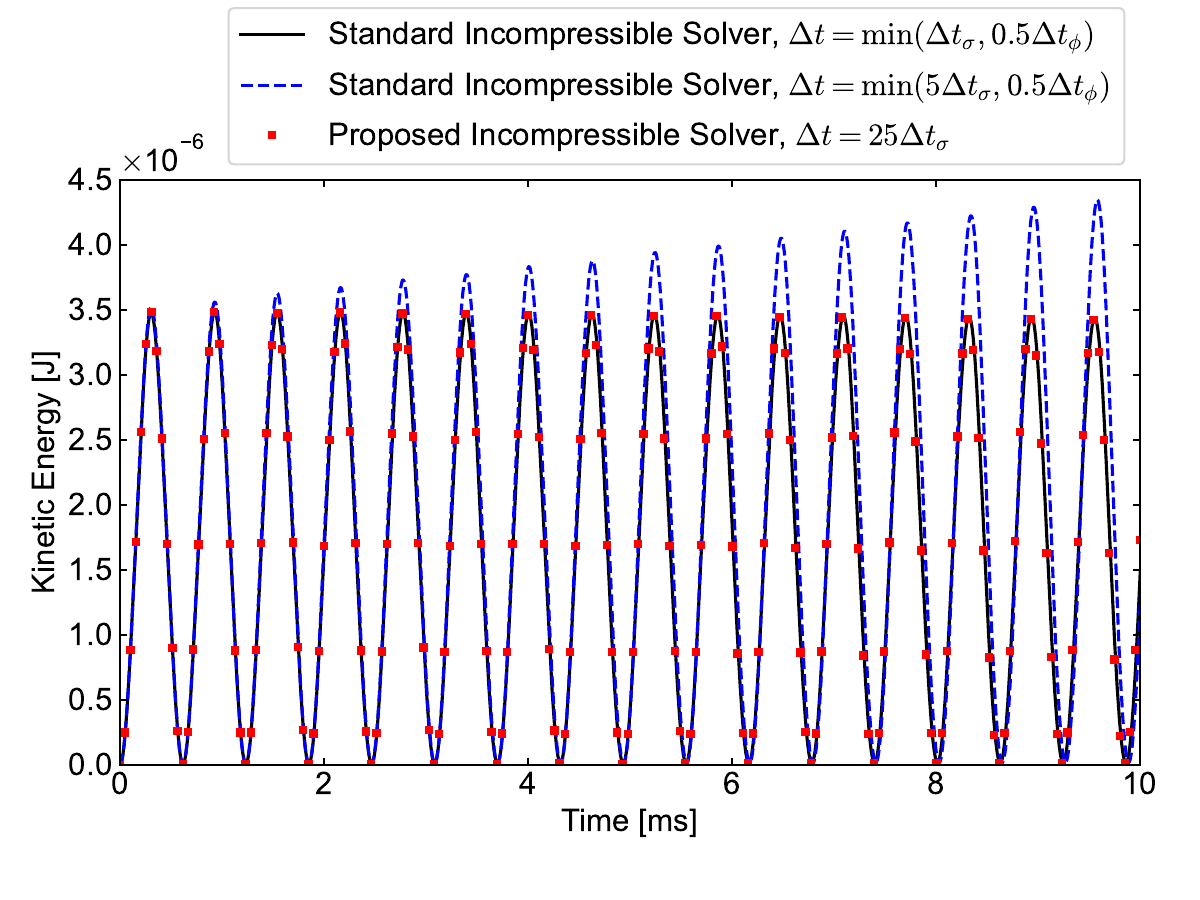}
    \caption{The evolution of kinetic energy for the inviscid 2D droplet oscillation in Section~\ref{sec:drop_oscillation}. The kinetic energy is computed as $\sum_{ij} \left( 0.5 \rho \|\bm{u}\|^2 \Delta x \Delta y \right)_{ij}$. The ``Standard Incompressible Solver'' denotes the solver in Section~\ref{sec:standard}. For the simulation with the proposed solver (red squares), the kinetic energy is plotted at every time step.}
    \label{fig:drop_oscillation:kinetic_energy}
\end{figure}
Figure~\ref{fig:drop_oscillation:kinetic_energy} shows the evolution of kinetic energy over the entire computational domain.
Since both phases are inviscid, the oscillation should not dissipate, and the successive peaks of kinetic energy should remain constant.
The standard incompressible solver described in Section~\ref{sec:standard} satisfies this requirement when the time-step size is sufficiently small, i.e., $\Delta t = \min(\Delta t_\sigma, 0.5 \Delta t_\phi)$.
However, using a larger time-step size, $\Delta t = \min(5 \Delta t_\sigma, 0.5 \Delta t_\phi)$, leads to a non-physical increase in kinetic energy.
This result demonstrates that the standard incompressible solver cannot significantly exceed the capillary time-step constraint, as is well known.
In contrast, the proposed method accurately reproduces the kinetic energy evolution even with $\Delta t = 25 \Delta t_\sigma$.
Notably, this time-step size is more than $4 \times$ the constraint of the phase-field method, $\min_t \Delta t_\phi$.  
The proposed method thus enables such a large time-step size without relying on artificially dissipative stabilization.
The absence of artificial dissipation is a key advantage of the proposed method compared with existing artificially dissipative surface tension treatments~\cite{Hysing2006, Raessi2009, DENNER201759, FAHSI2026106889}.

\subsection{2D Rising Bubble}
\label{sec:rising_bubble}

\paragraph{Objective}
Because the proposed method uses a weakly compressible solver in the substeps, readers may be concerned about velocity oscillations caused by acoustic waves, a well-known drawback of weakly compressible solvers~\cite{MATSUSHITA2019838, MATSUSHITA2021110605, YANG2021110113, MELVIN2025114195}. 
Here, we present 2D simulations of a rising bubble to demonstrate that the proposed method does not exhibit such oscillations.

\paragraph{Problem Setup}
\begin{figure}[!t]
    \centering
    \includegraphics[width=\linewidth]{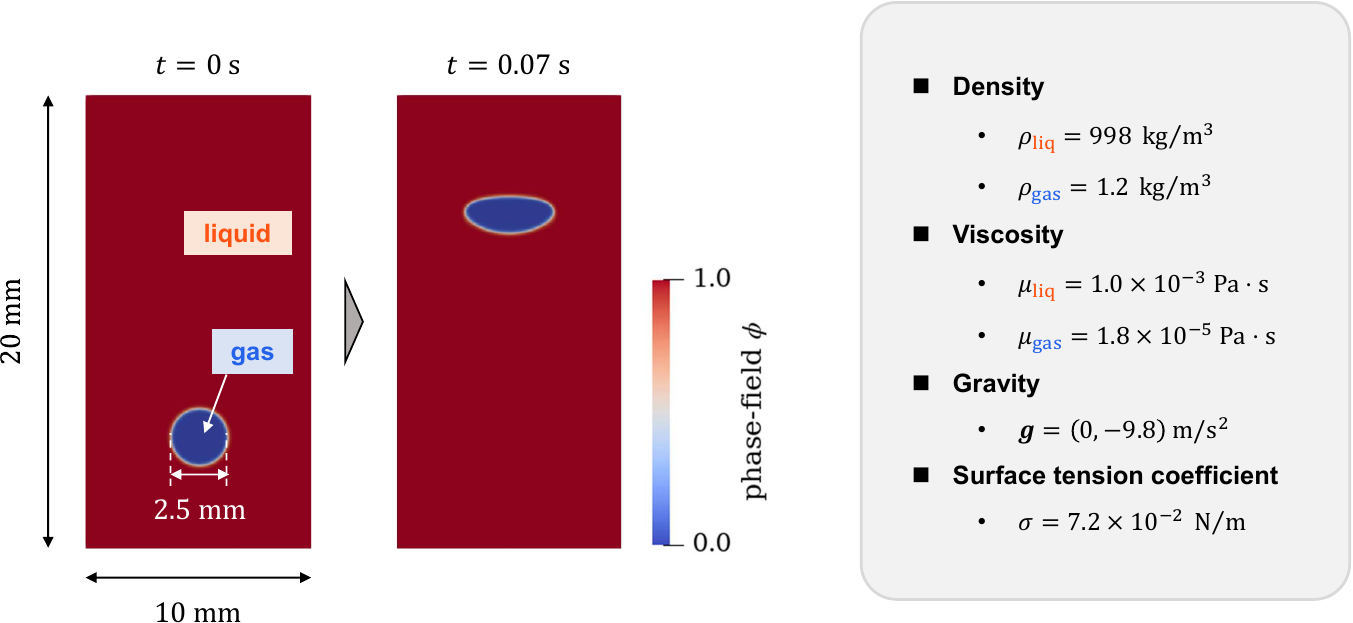}
    \caption{2D rising bubble test in Section~\ref{sec:rising_bubble}.}
    \label{fig:schematic_rising_bubble}
\end{figure}
Figure~\ref{fig:schematic_rising_bubble} illustrates the setup for this test.
A gas bubble with a diameter of $2.5~\mathrm{mm}$ is initially positioned at $(5, 5)~\mathrm{mm}$ within a domain of $[0,10]~\mathrm{mm} \times [0,20]~\mathrm{mm}$.
The liquid and gas densities are $\rho_\mathrm{liq} = 998~\mathrm{kg/m^3}$ and $\rho_\mathrm{gas} = 1.2~\mathrm{kg/m^3}$, respectively.
The viscosities are $\mu_\mathrm{liq} = 10^{-3}~\mathrm{Pa \cdot s}$ and $\mu_\mathrm{gas} = 1.8 \times 10^{-5}~\mathrm{Pa \cdot s}$. 
The surface tension coefficient is $\sigma = 7.2 \times 10^{-2}~\mathrm{N/m}$.
The gravitational acceleration is $\bm g = (0, -9.8)~\mathrm{m/s^2}$.
Initially, the velocity and pressure are set to $\bm u = 0$ and $p = 0$.
Free-slip boundary conditions ($u = 0$, $\partial v / \partial x = 0$) are applied at the left and right boundaries, whereas no-slip conditions are applied at the top and bottom boundaries.
The grid resolution is $256 \times 512$.
Under these settings, the dominant time-step constraints are the capillary ($\Delta t_\sigma \approx 8.1 \times 10^{-6}~\mathrm{s}$), the phase-field ($\min_t \Delta t_\phi \approx 2.4 \times 10^{-5}~\mathrm{s}$), and the viscosity ($\Delta t_\mu \approx 2.5 \times 10^{-5}~\mathrm{s}$).

\paragraph{Results (Rising Velocity)}
\begin{figure}[!t]
    \centering
    \includegraphics[width=0.6\linewidth]{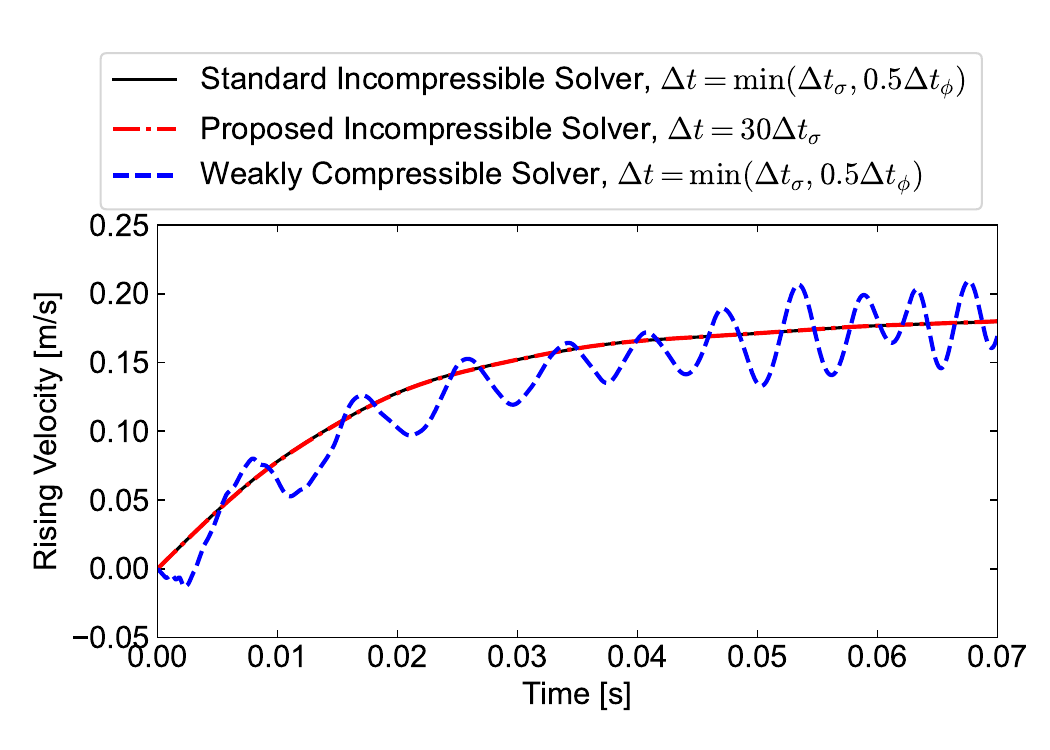}
    \caption{Bubble rising velocity for the test in Section~\ref{sec:rising_bubble}. The rising velocity is computed as $\sum_{ij} \left[(1-\phi)v\right]_{ij} \big/ \sum_{ij} (1-\phi)_{ij}$. The label ``Standard Incompressible Solver'' refers to the solver in Section~\ref{sec:standard}.
    The label ``Weakly Compressible Solver'' refers to the solver based on the evolving pressure projection method (see \cite{YANG2021110113, Yang2022} and stage 1 in Section~\ref{sec:subcycling}), with a time-step size of $\Delta t = \min(\Delta t_\sigma, 0.5\Delta t_\phi)$ and a sound speed of $c_s = 0.4\Delta h/\Delta t$.}
    \label{fig:rising_bubble:rise_velocity}
\end{figure}
Figure~\ref{fig:rising_bubble:rise_velocity} shows the bubble rising velocity.
The weakly compressible solver exhibits velocity oscillations due to large acoustic waves generated by the sudden pressure change at the start of the simulation.
Although increasing the sound speed $c_s$ effectively suppresses these oscillations~\cite{MATSUSHITA2019838}, it increases computational cost because the allowable time-step size is reduced ($\Delta t < O(\Delta x / c_s)$).
Another approach is to solve the pressure Poisson equation or to increase the number of iterations $N_\mathrm{EPP}$ only during the initial stage of the simulation~\cite{YANG2021110113, YAMASHITA2024113292}.
Although this approach also mitigates oscillations, it requires the timing of the sudden pressure change to be known in advance and is therefore not a general solution.
In contrast, the proposed incompressible solver does not exhibit such oscillations, even though the subcycling stage employs a weakly compressible solver.
The bubble rising velocity obtained with the proposed method agrees well with that of the standard incompressible solver, even when a large time-step size, $\Delta t = 30 \Delta t_\sigma$, is used.
Notably, this time-step size is more than $9 \times$ as large as both $\min_t \Delta t_\phi$ and $\Delta t_\mu$.
These results demonstrate that the proposed method can relax the capillary time-step constraint without the typical problem of weakly compressible solvers: oscillations caused by acoustic waves.

\subsection{Rayleigh--Plateau Instability}
\label{sec:rayleigh_plateau}

\paragraph{Objective}
We present simulations of the Rayleigh--Plateau instability.
The objective is to demonstrate the applicability and reduced computational cost of the proposed method for 3D simulations of capillary-driven instability.

\paragraph{Problem Setup}
\begin{figure}[!t]
    \centering
    \includegraphics[width=\linewidth]{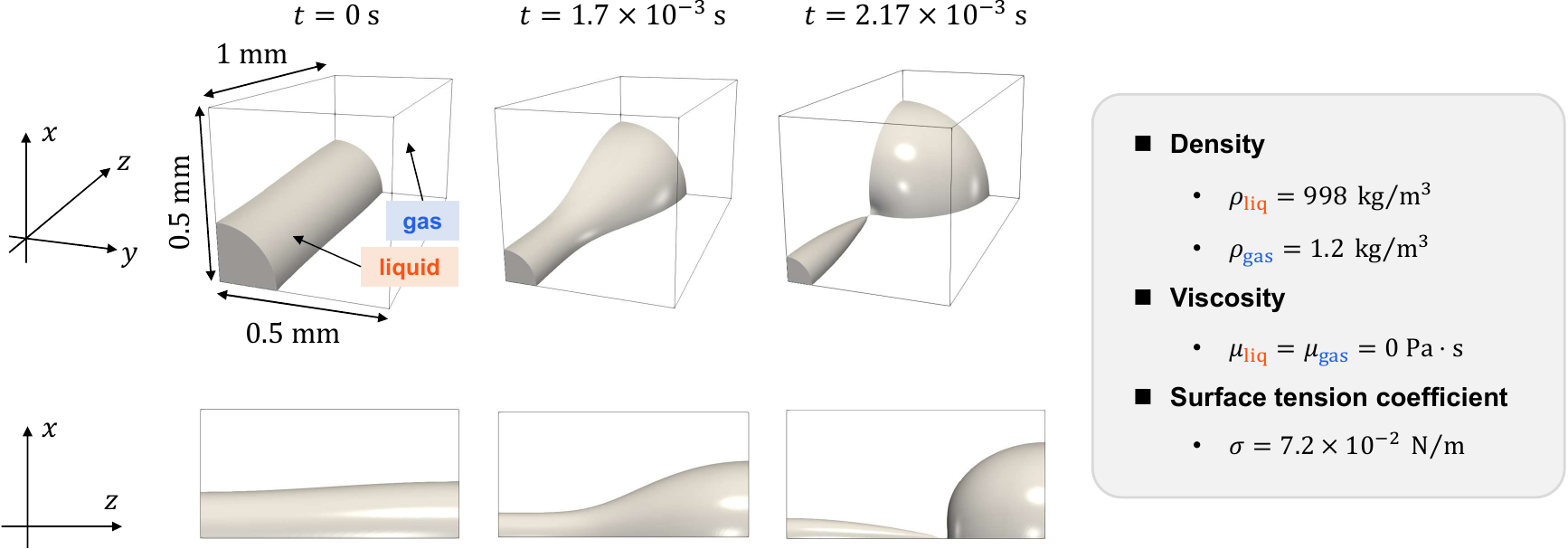}
    \caption{Simulation of the Rayleigh--Plateau instability in Section~\ref{sec:rayleigh_plateau}.}
    \label{fig:rayleigh_plateau:schematic}
\end{figure}
Figure~\ref{fig:rayleigh_plateau:schematic} shows the setup for this test.
A liquid column is initially placed within a 3D domain, $\left[0, 5 \times 10^{-4}\right]~\mathrm{m} \times \left[0, 5 \times 10^{-4}\right]~\mathrm{m} \times \left[-5 \times 10^{-4}, 5 \times 10^{-4}\right]~\mathrm{m}$.
The liquid column is a perturbed cylinder centered on the $z$-axis, with radius $r(z) = r_0 \left[1 + \alpha \sin(kz)\right]$, where $r_0 = 2 \times 10^{-4}~\mathrm{m}$, $\alpha = 0.1$, and $k = 1000\pi~\mathrm{rad/m}$.
The surrounding fluid is a gas.
The liquid and gas densities are $\rho_\mathrm{liq} = 998~\mathrm{kg/m^3}$ and $\rho_\mathrm{gas} = 1.2~\mathrm{kg/m^3}$, respectively.
Both phases are inviscid.
The surface tension coefficient is $\sigma = 7.2 \times 10^{-2}~\mathrm{N/m}$, and gravity is neglected.
Free-slip boundary conditions are applied at
$x = 0~\mathrm{m}$, $y = 0~\mathrm{m}$, $z = -5 \times 10^{-4}~\mathrm{m}$, and $z = 5 \times 10^{-4}~\mathrm{m}$, while outflow boundary conditions are applied at
$x = 5 \times 10^{-4}~\mathrm{m}$ and $y = 5 \times 10^{-4}~\mathrm{m}$.
The initial pressure is set to $p = 0$.
Following \cite{POPINET20095838}, the initial velocity field is set to
\begin{align}
    \bm{u} &= \nabla \Phi, \\
    \Phi &= \frac{\alpha r_0 c}{k}\frac{I_0(kr)}{I_1(kr_0)} \sin(kz), \\
    c &= \sqrt{\frac{\sigma k}{\rho r_0^2}\frac{I_1(kr_0)}{I_0(kr_0)}\left(1-k^2r_0^2\right)}
    = 1013.6,
\end{align}
where $I_0$ and $I_1$ are the modified Bessel functions.
The grid resolution is $200 \times 200 \times 400$.

\paragraph{Results (Minimum Radius and Pressure)}
\begin{figure}[!t]
    \centering
    \includegraphics[width=0.8 \linewidth]{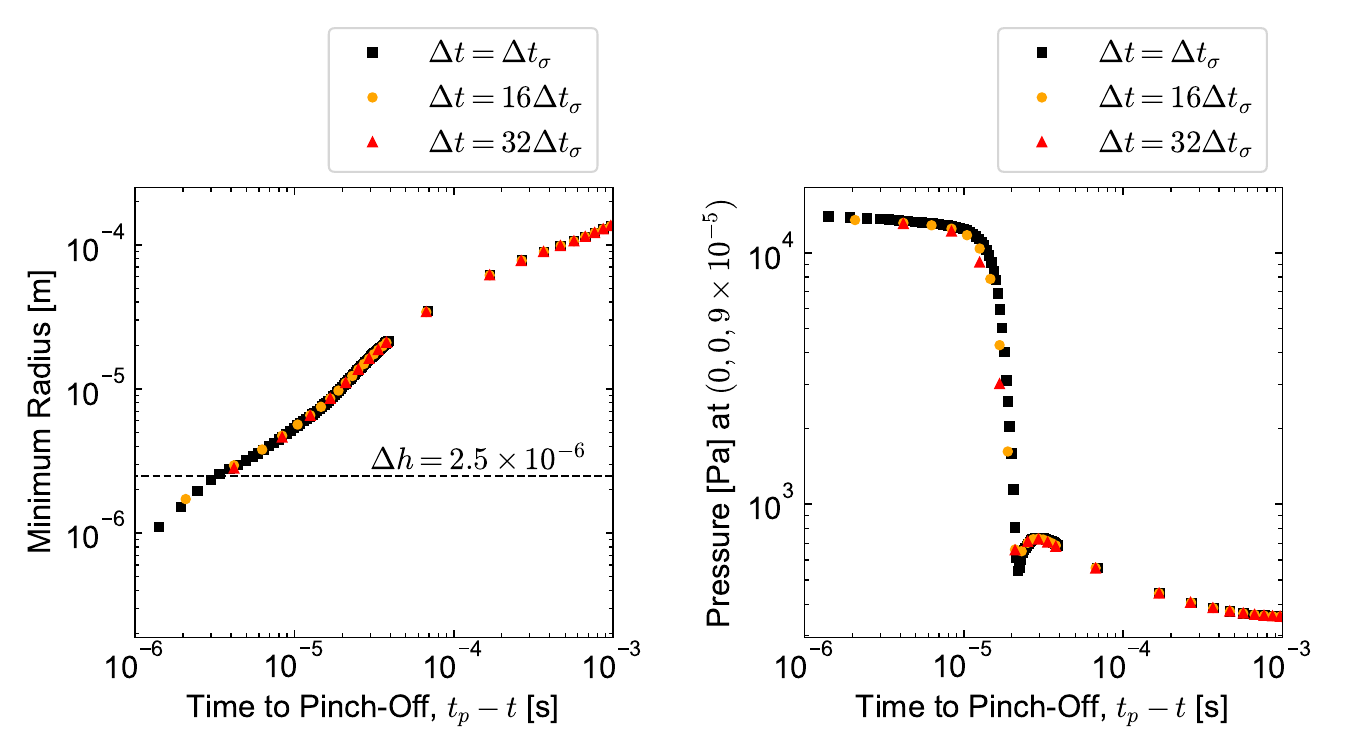}
    \caption{The minimum radius and the pressure near the neck in the simulation of the Rayleigh--Plateau instability in Section~\ref{sec:rayleigh_plateau}. See also Fig.~\ref{fig:rayleigh_plateau:neck_pressure} for the sudden pressure increase. The pinch-off time is $t_p = 2.169 \times 10^{-3}~\mathrm{s}$. The results are obtained using the proposed method.}
    \label{fig:rayleigh_plateau:vars_pinch_off}
\end{figure}
\begin{figure}[!t]
    \centering
    \includegraphics[width=\linewidth]{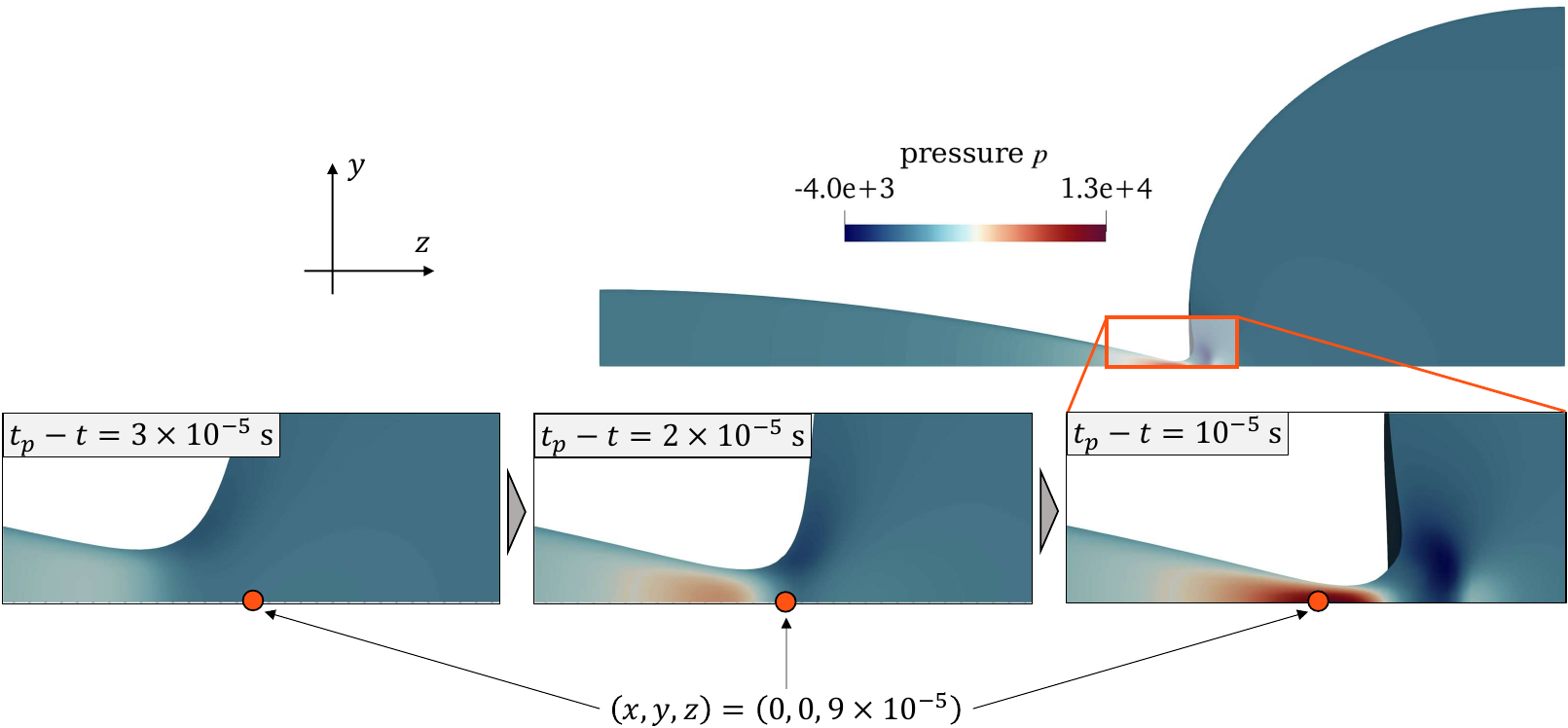}
    \caption{Sudden pressure increase near the neck in the simulation of the Rayleigh--Plateau instability in Section~\ref{sec:rayleigh_plateau}.}
    \label{fig:rayleigh_plateau:neck_pressure}
\end{figure}
Figure~\ref{fig:rayleigh_plateau:vars_pinch_off} shows the evolution of the minimum radius and the pressure near the neck.
The results are obtained using the proposed method with time-step sizes of $\Delta t = \Delta t_\sigma$, $16 \Delta t_\sigma$, and $32 \Delta t_\sigma$.
The pressure near the neck increases suddenly just before pinch-off.
This pressure rise is explained in Fig.~\ref{fig:rayleigh_plateau:neck_pressure}.
As the neck radius $r_\mathrm{neck}$ decreases, the pressure near the neck, $p \approx \sigma / r_\mathrm{neck}$, increases.
At the same time, the neck position moves in the positive $z$-direction.
The pressure at $(0, 0, 9 \times 10^{-5})$ suddenly increases when the point enters the high-pressure region.
For both the minimum radius and the pressure, the large time-step sizes ($16 \Delta t_\sigma$ and $32 \Delta t_\sigma$) yield results nearly identical to those obtained with $\Delta t = \Delta t_\sigma$, even when the neck radius is only a few grid cells.
These results demonstrate that the proposed method can exceed the capillary time-step constraint by at least 32 times in 3D simulations of Rayleigh--Plateau instability.

\paragraph{Results (Execution Time)}
\begin{figure}[!t]
    \centering
    \includegraphics[width=\linewidth]{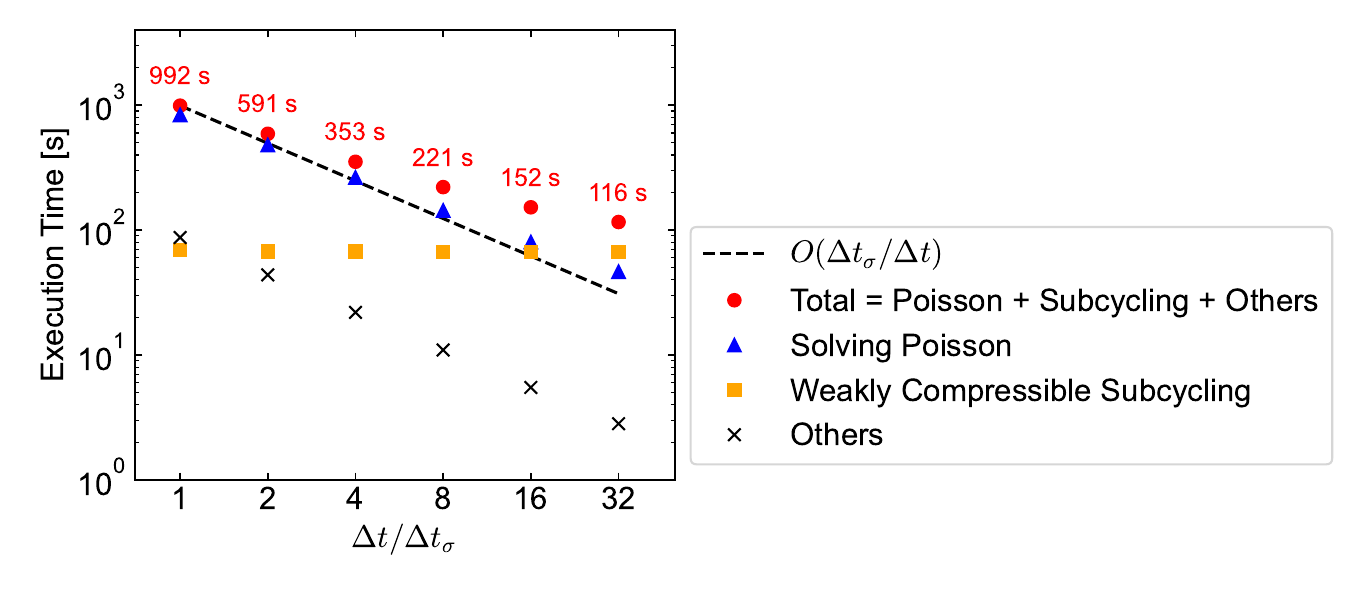}
    \caption{Execution time of the simulation of the Rayleigh--Plateau instability in Section~\ref{sec:rayleigh_plateau} for different time-step sizes. The execution time is measured up to $t = 3 \times 10^{-4}~\mathrm{s}$.}
    \label{fig:rayleigh_plateau:execution_time}
\end{figure}
Figure~\ref{fig:rayleigh_plateau:execution_time} shows the execution time of the proposed method for different time-step sizes.
We run the simulations on a single NVIDIA H100 GPU. 
The execution time decreases as the time-step size increases. 
Using $\Delta t = 32 \Delta t_\sigma$ yields an $8.6\times$ speedup over $\Delta t = \Delta t_\sigma$. 
The speedup is largest for small time-step sizes, where the pressure Poisson solver dominates the computational cost.
As the time-step size increases, the pressure Poisson solver accounts for a smaller fraction of the total execution time, resulting in a lower speedup.

\subsection{Two-Phase Flows in Porous Media}
\label{sec:porous2D}

\paragraph{Objective}
This section presents 2D simulations of two-phase flow in porous media.
The objective is to assess the applicability of the proposed method to one of the most important applications of surface-tension-dominated two-phase flow.

\paragraph{Problem Setup}
\begin{figure}[!t]
    \centering
    \includegraphics[width=\linewidth]{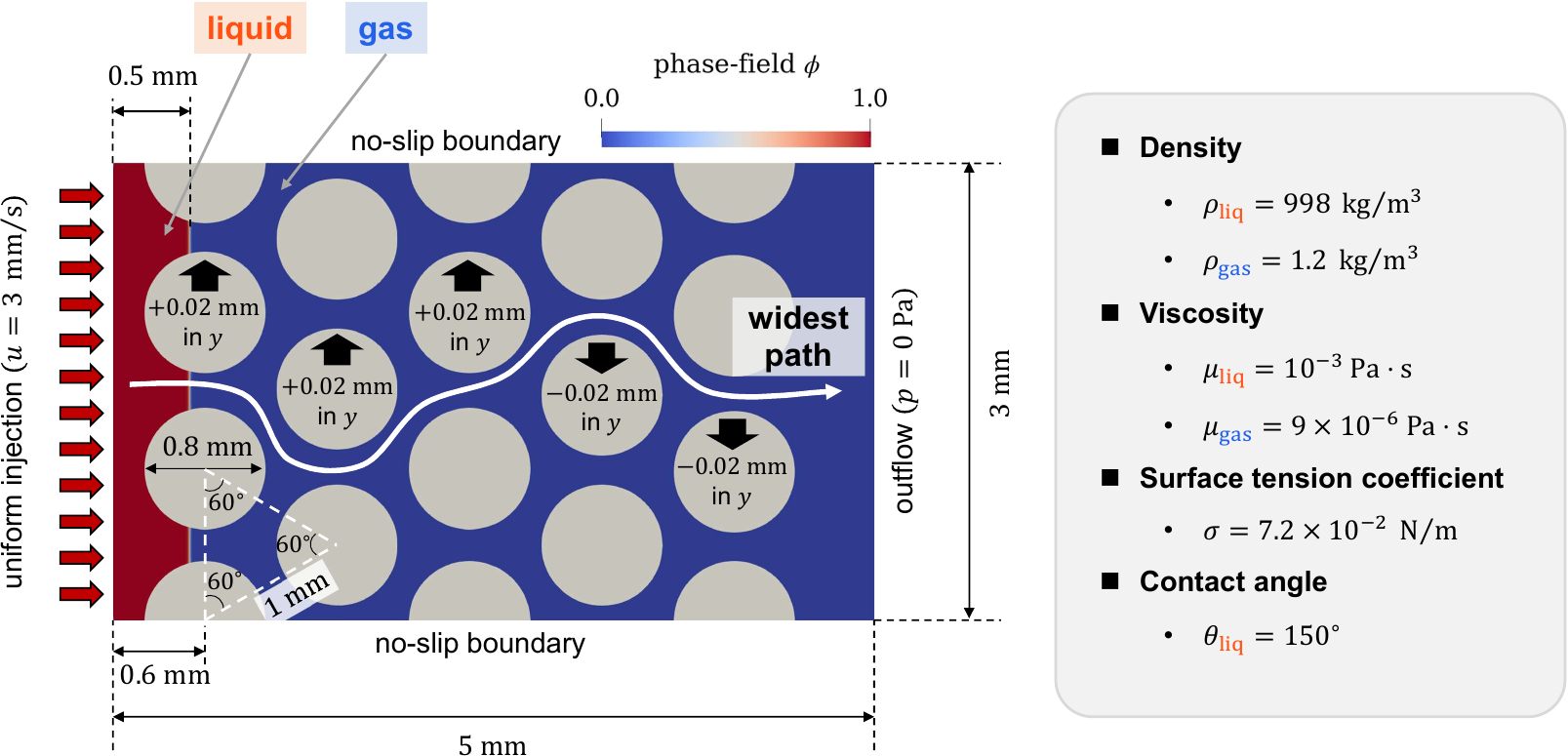}
    \caption{Initial settings for the 2D simulations of two-phase flow in porous media in Section~\ref{sec:porous2D}. The gray circles represent the solid objects that make up the porous medium. The objects are first arranged regularly with a spacing of $1~\mathrm{mm}$ between adjacent objects. Then, five objects are shifted by $0.02~\mathrm{mm}$ in the $y$ direction (black arrows), forming a single widest path (white arrow).}
    \label{fig:porous2D:settings}
\end{figure}
Figure~\ref{fig:porous2D:settings} shows the initial settings for the simulations.
Liquid is injected uniformly at $3~\mathrm{mm/s}$ at the left boundary, displacing the gas in the porous medium.
No-slip boundary conditions are applied to the top and bottom boundaries, whereas an outflow boundary condition with fixed pressure $p = 0$ is applied to the right boundary.
The porous medium is formed by arranging circular objects as follows: first, arrange them regularly; then, slightly shift five of them in the y-direction to create a single widest path.
See Fig.~\ref{fig:porous2D:settings} for more details.
The liquid and gas densities are $\rho_\mathrm{liq} = 998~\mathrm{kg/m^3}$ and $\rho_\mathrm{gas} = 1.2~\mathrm{kg/m^3}$, respectively.
The viscosities are $\mu_\mathrm{liq} = 10^{-3}~\mathrm{Pa \cdot s}$ and $\mu_\mathrm{gas} = 9 \times 10^{-6}~\mathrm{Pa \cdot s}$.
The surface tension coefficient is $\sigma = 7.2 \times 10^{-2}~\mathrm{N/m}$, and gravity is neglected.
The liquid--solid contact angle is $\theta_\mathrm{liq} = 150^\circ$; thus, the liquid is a non-wetting phase.
The grid resolution is $1250 \times 750$.
The objects are accounted for by the cut-cell method~\cite{MEYER20106300} for fluxes, in conjunction with the immersed boundary method~\cite{MITTAL20084825}.
For the contact angle, we extrapolate the phase-field variable into the objects, following~\cite{Yokoi2009contactangle}.

\paragraph{Results (Flow Pattern)}
\begin{figure}[!t]
    \centering
    \includegraphics[width=\linewidth]{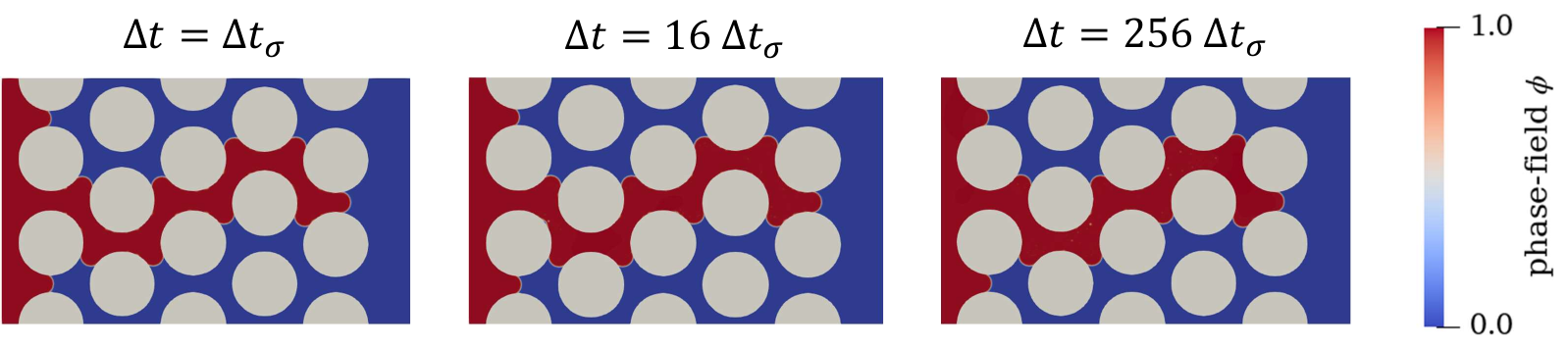}
    \caption{Flow patterns in porous media at $t = 0.18~\mathrm{s}$ obtained using the proposed method with three different time-step sizes (see Section~\ref{sec:porous2D}).}
    \label{fig:porous2D:capillary_fingering}
\end{figure}
Figure~\ref{fig:porous2D:capillary_fingering} shows the flow pattern for different time-step sizes.
All time-step sizes yield the same flow pattern, in which the liquid selectively invades the widest path.
This flow pattern is a well-known phenomenon called capillary fingering~\cite{Lu2019, Pavuluri2026}.
The interface preferentially invades wider throats because the interface curvature is lower there, requiring less liquid-phase pressure to overcome surface tension.
The proposed method can accurately reproduce this flow pattern while overcoming the capillary time-step constraint.

\paragraph{Results (Inlet--Outlet Pressure Difference)}
\begin{figure}[!t]
    \centering
    \includegraphics[width=0.5\linewidth]{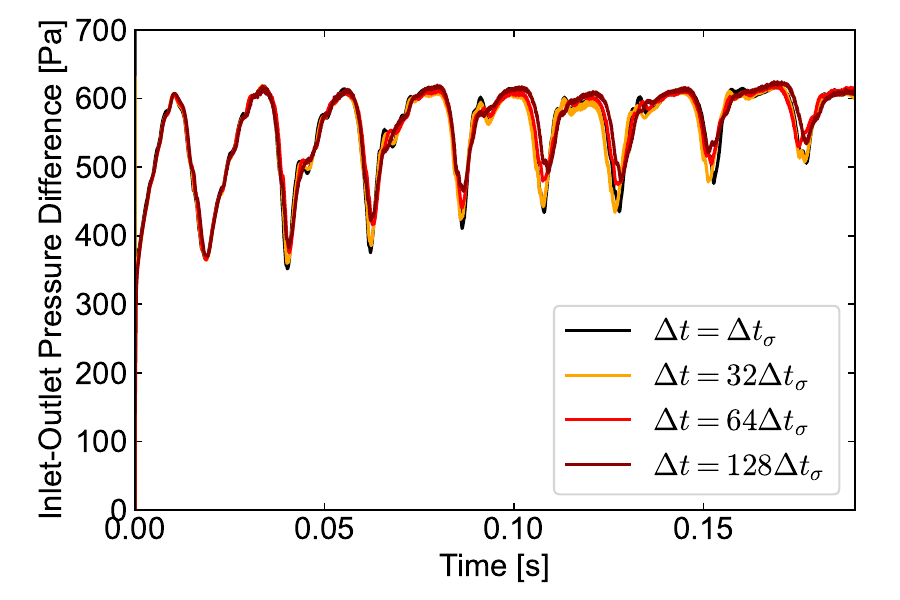}
    \caption{Inlet--outlet pressure difference in 2D porous media (see Section~\ref{sec:porous2D}). The simulations are performed using the proposed method with various time-step sizes. Figure~\ref{fig:porous2D:pressure_drop} provides a detailed view of the sudden pressure drop.}
    \label{fig:porous2D:pressure_diff}
\end{figure}
\begin{figure}[!t]
    \centering
    \includegraphics[width=\linewidth]{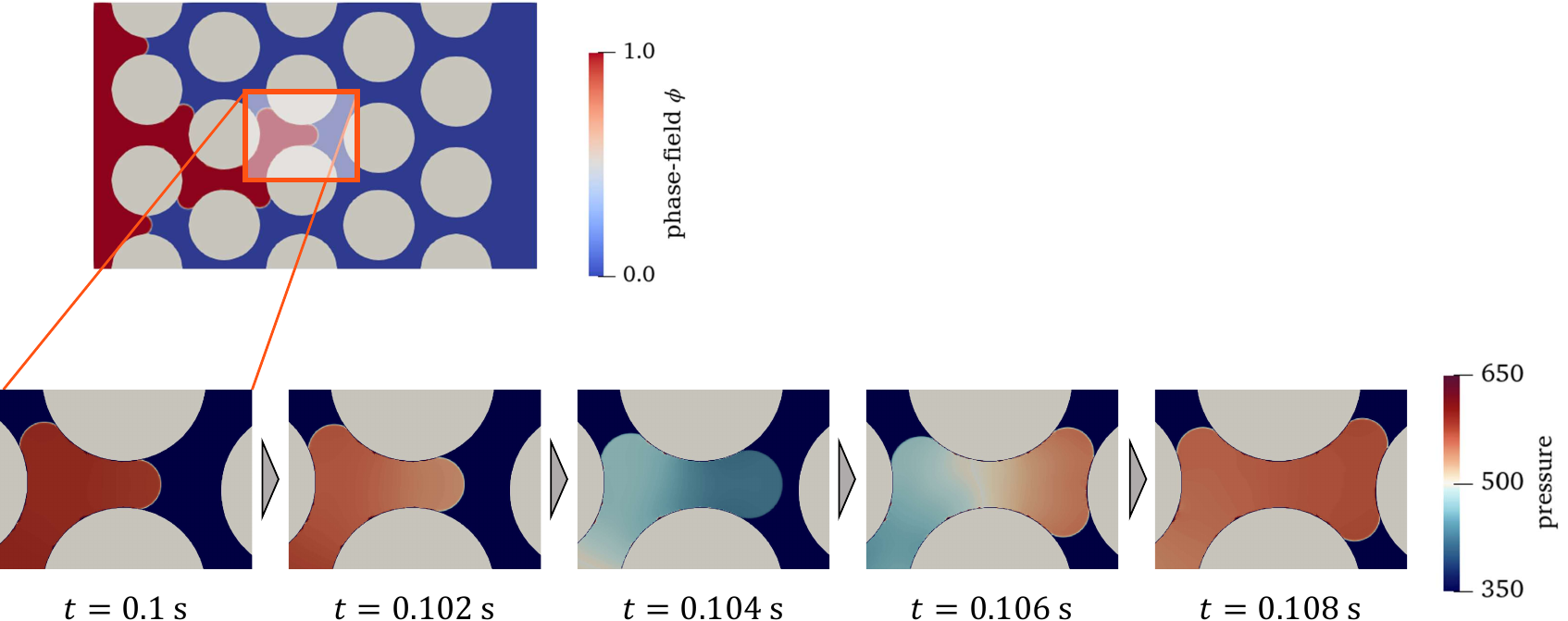}
    \caption{Sudden pressure drop when the liquid passes through the throat.}
    \label{fig:porous2D:pressure_drop}
\end{figure}
Figure~\ref{fig:porous2D:pressure_diff} shows the inlet--outlet pressure difference. 
The results exhibit sudden pressure drops at regular intervals. 
As shown in Fig.~\ref{fig:porous2D:pressure_drop}, these pressure drops occur when the interface passes through a throat and then suddenly expands into a pore. 
Simulations with time-step sizes up to $\Delta t = 32 \Delta t_\sigma$ yield pressure drops of nearly the same magnitude, whereas larger time-step sizes underestimate them. 
Therefore, the proposed method can exceed the capillary time-step constraint by up to $32\times$ in this test.

\paragraph{Results (Execution Time)}
\begin{figure}[!t]
    \centering
    \includegraphics[width=\linewidth]{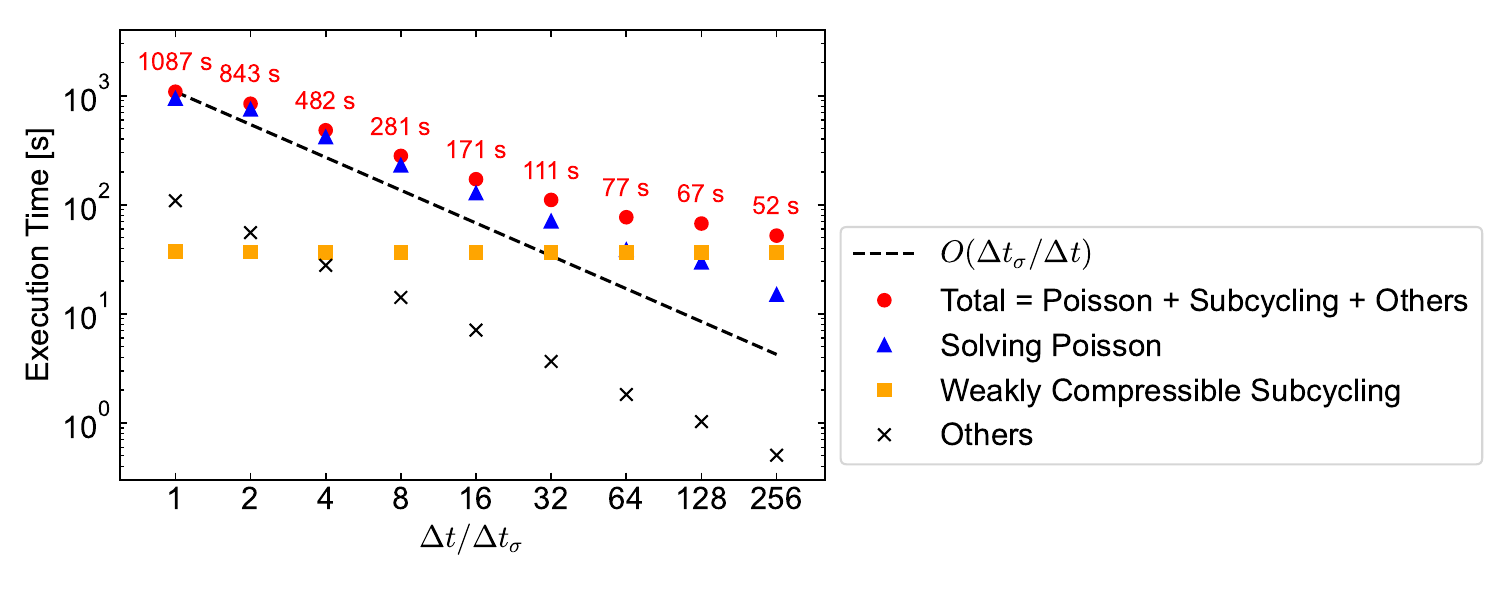}
    \caption{Execution time for the simulation of two-phase flow in porous media (Section~\ref{sec:porous2D}) with different time-step sizes. The execution time is measured up to $t = 0.005~\mathrm{s}$.}
    \label{fig:porous2D:execution_time}
\end{figure}
Figure~\ref{fig:porous2D:execution_time} shows the execution time for different time-step sizes. 
We run the simulations on a single NVIDIA H100 GPU. 
The execution time decreases as the time-step size increases. 
Using $\Delta t = 32 \Delta t_\sigma$ yields a $9.8\times$ speedup over $\Delta t = \Delta t_\sigma$. 
This speedup trend is similar to that observed in the previous test (Section~\ref{sec:rayleigh_plateau}). 
The speedup is largest for small time-step sizes, where the pressure Poisson solver dominates the computational cost. 
As the time-step size increases, the pressure Poisson solver accounts for a smaller fraction of the total execution time, resulting in a lower speedup.

\clearpage

\section{Conclusions}

\subsection{Summary}

In this study, we proposed an incompressible solver with weakly compressible subcycling (Fig.~\ref{fig:overview_subcycling}) to relax the capillary time-step constraint, thereby accelerating simulations of surface-tension-dominated incompressible two-phase flows (Section~\ref{sec:subcycling}).
The key advantage of the proposed method is that it relaxes the capillary time-step constraint without relying on artificially dissipative stabilization or requiring complex implementation, unlike previous mainstream time-implicit discretizations of surface tension.
Our strategy is to introduce lightweight substeps using a weakly compressible solver to assist the main incompressible solver.
These substeps enable the main incompressible solver to use accurately computed fluxes and surface tension force, even with large time-step sizes.
To the best of our knowledge, this is the first study to use a weakly compressible solver as an assistant to a main incompressible solver.

We demonstrated the applicability and efficiency of our method for practical 2D and 3D problems.
In the simulation of inviscid droplet oscillation (Section~\ref{sec:drop_oscillation}), we showed that the proposed method can use $\Delta t = 25 \Delta t_\sigma$ without relying on artificially dissipative stabilization, whereas the standard incompressible solver cannot use even $\Delta t = 5 \Delta t_\sigma$.
In the simulation of a rising bubble (Section~\ref{sec:rising_bubble}), we showed that the proposed method does not exhibit oscillations caused by acoustic waves, a well-known drawback of weakly compressible solvers.
In further practical problems, such as the Rayleigh--Plateau instability (Section~\ref{sec:rayleigh_plateau}) and two-phase flows in porous media (Section~\ref{sec:porous2D}), the proposed method achieves $8.6\times$ and $9.8\times$ speedups, respectively, with $\Delta t = 32 \Delta t_\sigma$ compared with simulations using $\Delta t = \Delta t_\sigma$.

\subsection{Limitations and Future Work}

\paragraph{Higher-Order Time Integration}
We use the first-order explicit Euler method for time integration.
This is because this study focuses on constructing a simple base framework and reducing computational cost.
We have not investigated extending our method to higher-order time integration methods, which may not be straightforward without increasing computational cost.

\paragraph{Compressibility}
In our method, we compute the fluxes using a weakly compressible solver.
Although we have not encountered any serious problems, the weak compressibility may adversely affect the phase-field variables, for example, by causing overshoots and undershoots.
Indeed, Fig.~\ref{fig:porous2D:capillary_fingering} shows that the simulation with $\Delta t = 256 \Delta t_\sigma$ produces tiny bubbles in the liquid phase.
Further improvement is needed to enhance the incompressibility of the computed fluxes.

\paragraph{Applicability to Broader Regimes}
This study focuses solely on surface-tension-dominated flows.
However, our method may also apply to other regimes.
Indeed, our method exceeds the time-step constraints imposed by the phase-field method and viscosity in simulations of droplet oscillation (Section~\ref{sec:drop_oscillation}) and a rising bubble (Section~\ref{sec:rising_bubble}).
Further research can clarify the applicability of our method not only to surface-tension-dominated cases but also to broader regimes.

\section*{Declaration of Generative AI and AI-Assisted Technologies in the Manuscript Preparation Process}

During the preparation of this work, the authors used ChatGPT and Grammarly solely to improve the readability of the English in this manuscript. Specifically, these tools were used to improve the readability of the original English text written by the authors and to translate the original Japanese text written by the authors into English. The authors reviewed all outputs and revised them as needed. The authors take full responsibility for the content of the manuscript.

\section*{Acknowledgements}

This work was supported by JSPS KAKENHI (Grant Numbers: JP26K21957, JP23K25024, JP25K07582, and JP22K14178), JST ASPIRE (Grant Number: 	JPMJAP2407), JST SPRING (Grant Number: JPMJSP2180), and Science Tokyo Support Program for Doctoral Students, funded by the Universities for International Research Excellence.
This study was carried out using the TSUBAME4.0 supercomputer at Institute of Science Tokyo.

\appendix

\clearpage

\section{Detailed Implementation of the Standard Incompressible Solver in Section~\ref{sec:standard}}
\label{sec:detailed_impl_standard}

We present the detailed implementation of the standard incompressible solver described in Section~\ref{sec:standard}.
For simplicity, we consider the 2D case.
Cell centers are denoted by the indices $(i, j)$, and cell faces are denoted by the indices $(i - \frac12, j)$ and $(i, j - \frac12)$.
The variables are the phase-field variable $\phi_{i, j}$, the cell-center velocity $\bm u_c = ((u_c)_{i, j}, (v_c)_{i, j})$, the cell-face velocity $\bm u_f = ((u_f)_{i - \frac12, j}, (v_f)_{i, j - \frac12})$, the pressure $p_{i, j}$, the density $\rho_{i, j}$, the viscosity $\mu_{i, j}$, the normal vector $\bm n_\phi = (({n_\phi}_x)_{i, j}, ({n_\phi}_y)_{i, j})$, and the curvature $\kappa_{i, j}$.
We use the notation $\langle q \rangle_{i - \frac12, j} = \frac12 (q_{i - 1, j} + q_{i, j})$ and $\langle q \rangle_{i, j - \frac12} = \frac12 (q_{i, j - 1} + q_{i, j})$.
The numerical procedure from time step $n$ to $n+1$ is as follows:
\begin{enumerate}
    \item Compute the volume flux $\bm F_\phi = \left(\left({F_\phi}_x\right)_{i - \frac12, j}, \left({F_\phi}_y\right)_{i, j - \frac12}\right)$ following~\cite{JAIN2022111529}:
    \begin{itemize}
        \item Compute $\psi_{i, j}^n = \epsilon \ln \left( \frac{\mathrm{clip}\left( \phi_{i, j}^n \right) + 10^{-100}}{1 - \mathrm{clip}\left( \phi_{i, j}^n \right) + 10^{-100}} \right)$, where $\mathrm{clip}\left( \phi_{i, j}^n \right) = \max\left(\min \left(\phi_{i, j}^n, 1 \right), 0\right)$.
        \item Compute $\Gamma = \left\| \bm u_f^n \right\|_\mathrm{max}$.
        \item Compute the volume flux $\bm F_\phi$:
        \begin{align}
            \left( {F_\phi}_x \right)_{i - \frac12, j} &= - \left( u_f^n \right)_{i - \frac12, j} \left\langle\phi^n\right\rangle_{i - \frac12, j} + \Gamma \left\{\epsilon \frac{\phi_{i, j}^n - \phi_{i - 1, j}^n}{\Delta x} - \frac14 \left[ 1 - \tanh^2 \left(\frac{\langle\psi^n\rangle_{i - \frac12, j}}{2 \epsilon}\right) \right] \left\langle {n_\phi^n}_x \right\rangle_{i - \frac12, j} \right\}, \\
            \left( {F_\phi}_y \right)_{i, j - \frac12} &= - \left( v_f^n \right)_{i, j - \frac12} \left\langle\phi^n\right\rangle_{i, j - \frac12} + \Gamma \left\{\epsilon \frac{\phi_{i, j}^n - \phi_{i, j - 1}^n}{\Delta y} - \frac14 \left[ 1 - \tanh^2 \left(\frac{\langle\psi^n\rangle_{i, j - \frac12}}{2 \epsilon}\right) \right] \left\langle {n_\phi^n}_y \right\rangle_{i, j - \frac12} \right\}.
        \end{align}
    \end{itemize}
    \item Compute the momentum fluxes $\bm F_{\rho u} = \left(\left({F_{\rho u}}_x\right)_{i - \frac12, j}, \left({F_{\rho u}}_y\right)_{i, j - \frac12}\right)$ and $\bm F_{\rho v} = \left(\left({F_{\rho v}}_x\right)_{i - \frac12, j}, \left({F_{\rho v}}_y\right)_{i, j - \frac12}\right)$:
    \begin{itemize}
        \item Compute the advective momentum fluxes $\bm F_{\rho u}^\mathrm{adv} = \left(\left({F_{\rho u}^\mathrm{adv}}_x\right)_{i - \frac12, j}, \left({F_{\rho u}^\mathrm{adv}}_y\right)_{i, j - \frac12}\right)$ and $\bm F_{\rho v}^\mathrm{adv} = \left(\left({F_{\rho v}^\mathrm{adv}}_x\right)_{i - \frac12, j}, \left({F_{\rho v}^\mathrm{adv}}_y\right)_{i, j - \frac12}\right)$ to be consistent with the volume flux $\bm F_\phi$~\cite{Yang2022}:
        \begin{itemize}
            \item Compute the mass flux $\bm F_\rho = \left( \left({F_\rho}_x\right)_{i - \frac12, j}, \left({F_\rho}_y\right)_{i, j - \frac12} \right)$:
            \begin{align}
                \left({F_\rho}_x\right)_{i - \frac12, j} &= -\rho_1 \left( u_f^n \right)_{i - \frac12, j} + \left( \rho_2 - \rho_1 \right) \left( {F_\phi}_x \right)_{i - \frac12, j},\\
                \left({F_\rho}_y\right)_{i, j - \frac12} &= -\rho_1 \left( v_f^n \right)_{i, j - \frac12} + \left( \rho_2 - \rho_1 \right) \left( {F_\phi}_y \right)_{i, j - \frac12}.
            \end{align}
            \item Compute the advective momentum fluxes $\bm F_{\rho u}^\mathrm{adv}$ and $\bm F_{\rho v}^\mathrm{adv}$:
            \begin{align}
                \left({F_{\rho u}^\mathrm{adv}}_x\right)_{i - \frac12, j} &= \left( {F_\rho}_x \right)_{i - \frac12, j} \mathrm{WENO}_{i - \frac12, j} \left( u_c^n \right), \\
                \left({F_{\rho u}^\mathrm{adv}}_y\right)_{i, j - \frac12} &= \left( {F_\rho}_y \right)_{i, j - \frac12} \mathrm{WENO}_{i, j - \frac12} \left( u_c^n \right), \\
                \left({F_{\rho v}^\mathrm{adv}}_x\right)_{i - \frac12, j} &= \left( {F_\rho}_x \right)_{i - \frac12, j} \mathrm{WENO}_{i - \frac12, j} \left( v_c^n \right), \\
                \left({F_{\rho v}^\mathrm{adv}}_y\right)_{i, j - \frac12} &= \left( {F_\rho}_y \right)_{i, j - \frac12} \mathrm{WENO}_{i, j - \frac12} \left( v_c^n \right),
            \end{align}
            where $\mathrm{WENO}_{i - \frac12, j} (q)$ and $\mathrm{WENO}_{i, j - \frac12} (q)$ denote the computation of $q_{i - \frac12, j}$ and $q_{i, j - \frac12}$ using the third-order WENO scheme~\cite{JIANG1996202}.
        \end{itemize}
        \item Compute the viscous stress $\tau = \mu \left[\nabla \bm u + \left(\nabla \bm u\right)^\top\right]$:
        \begin{align}
            \left(\tau_{xx}\right)_{i - \frac12, j} &= \left(2 \mu \frac{\partial u}{\partial x}\right)_{i - \frac12, j}^n = 2 \left\langle\mu^n\right\rangle_{i - \frac12, j} \frac{\left(u_c^n\right)_{i, j} - \left(u_c^n\right)_{i - 1, j}}{\Delta x}, \\
            \left(\tau_{xy}\right)_{i, j - \frac12} &= \left[\mu \left( \frac{\partial u}{\partial y} + \frac{\partial v}{\partial x} \right)\right]_{i, j - \frac12}^n = \left\langle\mu^n\right\rangle_{i, j - \frac12} \left(\frac{\left( u_c^n \right)_{i, j} - \left( u_c^n \right)_{i, j - 1}}{\Delta y} + \frac{\left\langle v_c^n \right\rangle_{i + 1, j - \frac12} - \left\langle v_c^n \right\rangle_{i - 1, j - \frac12}}{2 \Delta x}\right), \\
            \left(\tau_{yx}\right)_{i - \frac12, j} &= \left[\mu \left( \frac{\partial u}{\partial y} + \frac{\partial v}{\partial x} \right)\right]_{i - \frac12, j}^n = \left\langle\mu^n\right\rangle_{i - \frac12, j} \left(\frac{\left\langle u_c^n \right\rangle_{i - \frac12, j + 1} - \left\langle u_c^n \right\rangle_{i - \frac12, j - 1}}{2 \Delta y} + \frac{\left(v_c^n\right)_{i, j} - \left(v_c^n\right)_{i - 1, j}}{\Delta x}\right), \\
            \left(\tau_{yy}\right)_{i, j - \frac12} &= \left(2 \mu \frac{\partial v}{\partial y}\right)_{i, j - \frac12}^n = 2 \left\langle\mu^n\right\rangle_{i, j - \frac12} \frac{\left(v_c^n\right)_{i, j} - \left(v_c^n\right)_{i, j - 1}}{\Delta y}.
        \end{align}
        \item Compute the momentum fluxes $\bm F_{\rho u}$ and $\bm F_{\rho v}$:
        \begin{align}
            \left( {F_{\rho u}}_x \right)_{i - \frac12, j} = \left( {F_{\rho u}^\mathrm{adv}}_x \right)_{i - \frac12, j} + \left(\tau_{xx}\right)_{i - \frac12, j}, \\
            \left( {F_{\rho u}}_y \right)_{i, j - \frac12} = \left( {F_{\rho u}^\mathrm{adv}}_y \right)_{i, j - \frac12} + \left(\tau_{xy}\right)_{i, j - \frac12}, \\
            \left( {F_{\rho v}}_x \right)_{i - \frac12, j} = \left( {F_{\rho v}^\mathrm{adv}}_x \right)_{i - \frac12, j} + \left(\tau_{yx}\right)_{i - \frac12, j}, \\
            \left( {F_{\rho v}}_y \right)_{i, j - \frac12} = \left( {F_{\rho v}^\mathrm{adv}}_y \right)_{i, j - \frac12} + \left(\tau_{yy}\right)_{i, j - \frac12}.
        \end{align}
    \end{itemize}
    \item Compute the phase-field variable $\phi_{i, j}^{n + 1}$:
    \begin{equation}
        \phi_{i, j}^{n + 1} = \phi_{i, j}^n + \left( \frac{\left({F_\phi}_x\right)_{i + \frac12, j} - \left({F_\phi}_x\right)_{i - \frac12, j}}{\Delta x} + \frac{\left({F_\phi}_y\right)_{i, j + \frac12} - \left({F_\phi}_y\right)_{i, j - \frac12}}{\Delta y} \right) \Delta t.
    \end{equation}
    \item Compute the normal vector $\left(\bm n_\phi^{n + 1}\right)_{i, j}$, and the curvature $\kappa_{i, j}^{n + 1}$:
    \begin{itemize}
        \item Compute $\left(\nabla \psi^{n + 1}\right)_{i, j} = \left( \frac{\psi_{i + 1, j}^{n + 1} - \psi_{i - 1, j}^{n + 1}}{2 \Delta x}, \frac{\psi_{i, j + 1}^{n + 1} - \psi_{i, j - 1}^{n + 1}}{2 \Delta y} \right)$, where $\psi_{i, j}^{n + 1} = \epsilon \ln \left( \frac{\mathrm{clip}\left( \phi_{i, j}^{n + 1} \right) + 10^{-100}}{1 - \mathrm{clip}\left( \phi_{i, j}^{n + 1} \right) + 10^{-100}} \right)$.
        \item Compute the normal vector $(\bm n_\phi^{n + 1})_{i, j} = \frac{\left(\nabla \psi^{n + 1}\right)_{i, j}}{\left\|\left( \nabla \psi^{n + 1} \right)_{i, j}\right\| + 10^{-100}}$.
        \item Compute the curvature $\kappa_{i, j}^{n + 1} = - \left(\nabla \cdot \bm n_\phi^{n + 1}\right)_{i, j} = -\left( \frac{\left( {n_\phi^{n + 1}}_x \right)_{i + 1, j} - \left( {n_\phi^{n + 1}}_x \right)_{i - 1, j}}{2 \Delta x}, \frac{\left( {n_\phi^{n + 1}}_y \right)_{i, j + 1} - \left( {n_\phi^{n + 1}}_y \right)_{i, j - 1}}{2 \Delta y} \right)$.
    \end{itemize}
    \item Compute the density $\rho_{i, j}^{n + 1} = \rho_1 \left(1 - \phi_{i, j}^{n + 1}\right) + \rho_2 \phi_{i, j}^{n + 1}$ and the viscosity $\mu_{i, j}^{n + 1} = \mu_1 \left(1 - \phi_{i, j}^{n + 1}\right) + \mu_2 \phi_{i, j}^{n + 1}$.
    \item Compute the intermediate cell-center velocity $\left( \bm u_c^* \right)_{i, j} = \left( \left( u_c^* \right)_{i, j}, \left( v_c^* \right)_{i, j} \right)$:
    \begin{align}
        \left( u_c^* \right)_{i, j} &= \frac{\rho_{i, j}^n \left( u_c^n \right)_{i, j} + \left( \frac{\left( {F_{\rho u}}_x \right)_{i + \frac12, j} - \left( {F_{\rho u}}_x \right)_{i - \frac12, j}}{\Delta x} + \frac{\left( {F_{\rho u}}_y \right)_{i, j + \frac12} - \left( {F_{\rho u}}_y \right)_{i, j - \frac12}}{\Delta y} \right) \Delta t}{\rho_{i, j}^{n + 1}}, \\
        \left( v_c^* \right)_{i, j} &= \frac{\rho_{i, j}^n \left( v_c^n \right)_{i, j} + \left( \frac{\left( {F_{\rho v}}_x \right)_{i + \frac12, j} - \left( {F_{\rho v}}_x \right)_{i - \frac12, j}}{\Delta x} + \frac{\left( {F_{\rho v}}_y \right)_{i, j + \frac12} - \left( {F_{\rho v}}_y \right)_{i, j - \frac12}}{\Delta y} \right) \Delta t}{\rho_{i, j}^{n + 1}}.
    \end{align}
    \item Compute the intermediate cell-face velocity $\bm u_f^* = \left( \left( u_f^* \right)_{i - \frac12, j}, \left( v_f^* \right)_{i, j - \frac12} \right) = \left( \left\langle u_c^* \right\rangle_{i - \frac12, j}, \left\langle v_c^* \right\rangle_{i, j - \frac12} \right)$.
    \item Update the intermediate velocities $\bm u_f^*$ and $\bm u_c^*$ using the surface tension force $\bm f_\sigma$:
    \begin{itemize}
        \item Compute the surface tension force $\bm f_\sigma = \left(\left( {f_\sigma}_x \right)_{i - \frac12, j}, \left( {f_\sigma}_y \right)_{i, j - \frac12}\right)$:
        \begin{align}
            \left( {f_\sigma}_x \right)_{i - \frac12, j} &= 6 \left\langle \phi^{n + 1} \right\rangle_{i - \frac12, j} \left( 1 - \left\langle \phi^{n + 1} \right\rangle_{i - \frac12, j} \right) \sigma \left\langle \kappa^{n + 1} \right\rangle_{i - \frac12, j} \frac{\phi_{i, j}^{n + 1} - \phi_{i - 1, j}^{n + 1}}{\Delta x}, \\
            \left( {f_\sigma}_y \right)_{i, j - \frac12} &= 6 \left\langle \phi^{n + 1} \right\rangle_{i, j - \frac12} \left( 1 - \left\langle \phi^{n + 1} \right\rangle_{i, j - \frac12} \right) \sigma \left\langle \kappa^{n + 1} \right\rangle_{i, j - \frac12} \frac{\phi_{i, j}^{n + 1} - \phi_{i, j - 1}^{n + 1}}{\Delta y}.
        \end{align}
        \item Update the intermediate velocities:
        \begin{align}
            \bm u_f^* &\leftarrow \left( \left( u_f^* \right)_{i - \frac12, j} + \frac{\left( {f_\sigma}_x \right)_{i - \frac12, j} \Delta t}{\left\langle \rho^{n + 1} \right\rangle_{i - \frac12, j}}, \left( v_f^* \right)_{i, j - \frac12} + \frac{\left( {f_\sigma}_y \right)_{i, j - \frac12} \Delta t}{\left\langle \rho^{n + 1} \right\rangle_{i, j - \frac12}} \right), \\
            \bm u_c^* &\leftarrow \left( \left( u_c^* \right)_{i, j} + \frac12 \left( \frac{\left( {f_\sigma}_x \right)_{i - \frac12, j} \Delta t}{\left\langle \rho^{n + 1} \right\rangle_{i - \frac12, j}} + \frac{\left( {f_\sigma}_x \right)_{i + \frac12, j} \Delta t}{\left\langle \rho^{n + 1} \right\rangle_{i + \frac12, j}} \right), \left( v_c^* \right)_{i, j} + \frac12 \left( \frac{\left( {f_\sigma}_y \right)_{i, j - \frac12} \Delta t}{\left\langle \rho^{n + 1} \right\rangle_{i, j - \frac12}} + \frac{\left( {f_\sigma}_y \right)_{i, j + \frac12} \Delta t}{\left\langle \rho^{n + 1} \right\rangle_{i, j + \frac12}} \right) \right).
        \end{align}
    \end{itemize}
    \item Solve the pressure Poisson equation:
    \begin{itemize}
        \item Compute the velocity divergence $\left( \nabla \cdot \bm u_f^* \right)_{i, j} = \frac{\left( u_f^* \right)_{i + \frac12, j} - \left( u_f^* \right)_{i - \frac12, j}}{\Delta x} + \frac{\left( v_f^* \right)_{i, j + \frac12} - \left( v_f^* \right)_{i, j - \frac12}}{\Delta y}$.
        \item Solve the pressure Poisson equation:
        \begin{equation}
            \frac{\frac{1}{\left\langle \rho^{n + 1} \right\rangle_{i + \frac12, j}} \frac{p_{i + 1, j}^{n + 1} - p_{i, j}^{n + 1}}{\Delta x} - \frac{1}{\left\langle \rho^{n + 1} \right\rangle_{i - \frac12, j}} \frac{p_{i, j}^{n + 1} - p_{i - 1, j}^{n + 1}}{\Delta x}}{\Delta x} +
            \frac{\frac{1}{\left\langle \rho^{n + 1} \right\rangle_{i, j + \frac12}} \frac{p_{i, j + 1}^{n + 1} - p_{i, j}^{n + 1}}{\Delta y} - \frac{1}{\left\langle \rho^{n + 1} \right\rangle_{i, j - \frac12}} \frac{p_{i, j}^{n + 1} - p_{i, j - 1}^{n + 1}}{\Delta y}}{\Delta y} =
            \left( \nabla \cdot \bm u_f^* \right)_{i, j}.
        \end{equation}
        We solve this equation as a linear system by first applying diagonal scaling and then using the FlexGMRES solver with the PFMG preconditioner provided by the \textit{hypre} library~\cite{hypre, Falgout2002}, with a convergence tolerance of $10^{-7}$.
    \end{itemize}
    \item Compute the velocities at time step $n+1$:
    \begin{align}
        \bm u_f^{n + 1} &= \left( \left( u_f^* \right)_{i - \frac12, j} - \frac{\frac{p_{i, j}^{n + 1} - p_{i - 1, j}^{n + 1}}{\Delta x}}{\left\langle \rho^{n + 1} \right\rangle_{i - \frac12, j}} \Delta t, \left( v_f^* \right)_{i, j - \frac12} - \frac{\frac{p_{i, j}^{n + 1} - p_{i, j - 1}^{n + 1}}{\Delta y}}{\left\langle \rho^{n + 1} \right\rangle_{i, j - \frac12}} \Delta t \right), \\
        \bm u_c^{n + 1} &= \left( 
            \left( u_c^* \right)_{i, j} - \frac12 \left(\frac{\frac{p_{i, j}^{n + 1} - p_{i - 1, j}^{n + 1}}{\Delta x}}{\left\langle \rho^{n + 1} \right\rangle_{i - \frac12, j}} + \frac{\frac{p_{i + 1, j}^{n + 1} - p_{i, j}^{n + 1}}{\Delta x}}{\left\langle \rho^{n + 1} \right\rangle_{i + \frac12, j}} \right) \Delta t,
            \left( v_c^* \right)_{i, j} - \frac12 \left(\frac{\frac{p_{i, j}^{n + 1} - p_{i, j - 1}^{n + 1}}{\Delta y}}{\left\langle \rho^{n + 1} \right\rangle_{i, j - \frac12}} + \frac{\frac{p_{i, j + 1}^{n + 1} - p_{i, j}^{n + 1}}{\Delta y}}{\left\langle \rho^{n + 1} \right\rangle_{i, j + \frac12}} \right) \Delta t,
            \right).
    \end{align}
\end{enumerate}

\bibliographystyle{elsarticle-num}
\bibliography{reference}

\end{document}